\documentclass[a4paper,11pt]{article}
\usepackage{jheppub}           

\usepackage[T1]{fontenc}       
\usepackage[utf8]{inputenc}    
\usepackage{microtype}
\usepackage{csquotes}

\usepackage{amsmath,amssymb,mathtools,bm}
\usepackage{microtype}         
\usepackage{graphicx,booktabs} 
\usepackage{xcolor}            
\usepackage{slashed}
\usepackage[nameinlink,capitalise]{cleveref} 
\usepackage{tikz}
\usepackage[compat=1.1.0]{tikz-feynhand}
\usetikzlibrary{decorations.pathmorphing}
\usetikzlibrary{arrows.meta,positioning,calc}
\usepackage{adjustbox}
\newlength{\boxw}
\tikzset{
  box/.style={
    draw, rounded corners, align=center,
    text width=\boxw, minimum height=12mm,
    inner sep=3.5pt, line width=0.8pt
  },
  arr/.style={-{Latex[length=2.6mm]}, line width=0.9pt, shorten >=1.2pt, shorten <=1.2pt}
}
\newcommand{\makeports}[1]{%
  \coordinate (#1N) at ($ (#1.north)!0.08!(#1.center) $);
  \coordinate (#1S) at ($ (#1.south)!0.08!(#1.center) $);
  \coordinate (#1E) at ($ (#1.east)!0.08!(#1.center) $);
  \coordinate (#1W) at ($ (#1.west)!0.08!(#1.center) $);
}

\DeclarePairedDelimiter{\pqty}{(}{)}     
\DeclarePairedDelimiter{\bqty}{[}{]}     
\DeclarePairedDelimiter{\Bqty}{\{}{\}}   
\DeclarePairedDelimiter{\abs}{\lvert}{\rvert}
\DeclarePairedDelimiter{\norm}{\lVert}{\rVert}

\newcommand{\dd}{\mathop{}\!\mathrm{d}}
\newcommand{\dv}[3][]{\frac{\mathrm{d}^{#1} #2}{\mathrm{d} #3^{#1}}}
\newcommand{\pdv}[3][]{\frac{\partial^{#1} #2}{\partial #3^{#1}}}

\DeclareMathOperator{\tr}{tr}

\DeclareMathOperator{\Disc}{Disc}
\DeclareMathOperator{\diag}{diag}

\let\Im\relax
\DeclareMathOperator{\Im}{Im}

\newcommand{\cO}{\mathcal{O}}

\newcommand{\YB}{Y_B}

\newcommand{\gstar}{g_{\star}}
\newcommand{\Besselk}{\mathrm{K}}

\newcommand{\CNB}{C_{NB}}      
\newcommand{\ONB}{\mathcal{O}_{NB}}      

\newcommand{\beq}{\begin{equation}}
\newcommand{\eeq}{\end{equation}}

\numberwithin{equation}{section}

\allowdisplaybreaks

\preprint{RESCEU-17/25}

\title{Electromagnetic Leptogenesis --- an EFT-Consistent Analysis
via Wilson Coefficients. Part I. Low-Scale, Non-Resonant Regime}

\author[a]{Rin Takada}
\affiliation[a]{Research Center for the Early Universe (RESCEU), Graduate School of Science, The University of Tokyo, 7-3-1 Hongo, Bunkyo, Tokyo 113-0033, Japan}
\emailAdd{takada-rin@resceu.s.u-tokyo.ac.jp}

\abstract{
We analyse electromagnetic leptogenesis within the framework of an effective field theory, where the dynamics is governed by the gauge-invariant dipole operator $\ONB=(\bar{L}\sigma^{\mu\nu}N)\tilde{H}B_{\mu\nu}$. The Wilson coefficient $\CNB$ is matched at one loop and evolved to the electroweak scale via renormalisation-group (RG) running. After electroweak symmetry breaking, we compute flavour-dependent decay widths and CP asymmetries for the two-body modes $N\to\nu\gamma$ and $N\to\nu Z$, and solve the fully flavoured Boltzmann equations. In the $N_1$-dominated regime, the freeze-out baryon asymmetry is $Y_B^{\rm FO}\lesssim 10^{-17}$, far below the observed value $Y_B^{\rm obs}\simeq 8.7\times 10^{-11}$. The suppression is structural: gauge invariance forces a Higgs insertion; therefore the dipole coupling scales as $\mu\propto v/M_{\Psi}^2$, while the matched Wilson coefficient $\CNB$ is loop-generated and further suppressed by RG running. We note that in the quasi-degenerate limit the self-energy resonance can be operative and suggest a plausible path to $Y_B^{\rm obs}$.
}

\keywords{Baryo-and Leptogenesis, Early Universe Particle Physics, Sterile or Heavy Neutrinos}
\arxivnumber{2509.07698}

\begin{document}
\maketitle
\flushbottom

\section{Introduction}
\label{sec:1}

The observed baryon asymmetry of the Universe (BAU) points to dynamics that satisfy Sakharov's three conditions~\cite{Sakharov:1967dj}: (i) baryon-number violation, (ii) violation of $C$ and $CP$ symmetries, and (iii) a departure from thermal equilibrium. In the Standard Model (SM), anomalous electroweak transitions---sphalerons~\cite{PhysRevD.28.2019, PhysRevD.30.2212}---violate $B+L$ while conserving $B-L$, thereby enabling the conversion of a lepton asymmetry into a baryon asymmetry. Fukugita and Yanagida proposed leptogenesis~\cite{FUKUGITA198645}, which provides a robust mechanism for producing the BAU. In leptogenesis, the out-of-equilibrium, CP-violating decays of heavy (typically Majorana) right-handed neutrinos generate a lepton asymmetry, which sphalerons partially convert into a baryon asymmetry. Beyond its conceptual simplicity, leptogenesis naturally correlates with frameworks that account for light-neutrino masses (e.g. the type-I seesaw~\cite{Minkowski:1977sc, Yanagida:1979as, Gell-Mann:1979vob, PhysRevLett.44.912}), and has matured into a quantitatively predictive framework; see standard reviews~\cite{BUCHMULLER2005305, DAVIDSON2008105}. 

Bell--Kayser--Law~\cite{PhysRevD.78.085024} first proposed the possibility of electromagnetic leptogenesis (EMLG). They showed that an electromagnetic dipole operator induces the radiative two-body decays $N\to\nu+\gamma/Z$ and thus provides the CP-violating source when loop corrections are taken into account. Moreover, resonant EMLG at the TeV scale~\cite{Choudhury:2011gbi} was explored in the context of resonant leptogenesis~\cite{PILAFTSIS2004303, DEV2016268}. In the previous analysis~\cite{PhysRevD.78.085024}, the relevant couplings are effectively treated as arbitrary inputs. By contrast, in this work we compute the Wilson coefficient of the gauge-invariant dipole operator $\ONB=(\bar{L}\sigma^{\mu\nu}P_RN)\tilde{H}B_{\mu\nu}$ at one loop within the effective field theory (EFT) framework.

This work presents a first-principles EFT analysis of electromagnetic leptogenesis in the electroweak-broken phase, formulated within a gauge-invariant effective field theory (EFT) and defined at the level of Wilson coefficients. We derive the one-loop Wilson coefficient $\CNB$ from a renormalisable UV completion, evolve it with the renormalisation group down to an electroweak reference scale, and---after the electroweak symmetry breaking (EWSB)---map it onto the dipole coupling,
\beq
\label{eq:1.1}
\mu=\dfrac{v\cos\theta_{\rm W}}{\sqrt{2}\,M_{\Psi}^2}\,
\CNB(\mu_{\rm ref}),
\eeq
which then enters the decay widths $\varGamma$, the CP asymmetries $\varepsilon_{\alpha i}$. In the broken phase, the two-body channels are open. We then solve the three-flavour Boltzmann equations including spectator effects~\cite{Nardi_2006, Abada_2006, DAVIDSON2008105, ANTUSCH2012180} in sec.~\ref{sec:5} and present the numerical results in sec.~\ref{sec:6}, with their implications discussed in sec.~\ref{sec:7}.

From the one-loop matching (sec.~\ref{sec:2}) the Wilson coefficient is schematically
\beq
\label{eq:1.2}
\CNB\sim\dfrac{1}{16\pi^2}f(r,\xi_j)\times(\text{couplings}),
\eeq
where $r$ and $\xi_j$ denote heavy-mass ratios that are kept fixed in our scans, and $f(r,\xi_j)=\cO(1)$ is a dimensionless loop function. Thus, after factoring out the couplings and for fixed $\{r,\xi_j\}$,
\beq
\label{eq:1.3}
\mu\propto\dfrac{1}{16\pi^2}\cdot\dfrac{v}{M_{\Psi}^2}.
\eeq
This double suppression---by the loop factor $1/(16\pi^2)$ and by the inverse heavy-mass dependence $v/M_{\Psi}^2$---feeds into both the production source $\varepsilon$ and the washout in the flavour-resolved Boltzmann system (sec.~\ref{sec:5}). Numerically, for representative non-resonant benchmarks with hierarchical heavy spectra, this scaling renders the dipole couplings and hence the CP source very small; the quantitative impact is assessed in secs.~\ref{sec:6}--\ref{sec:7}. 

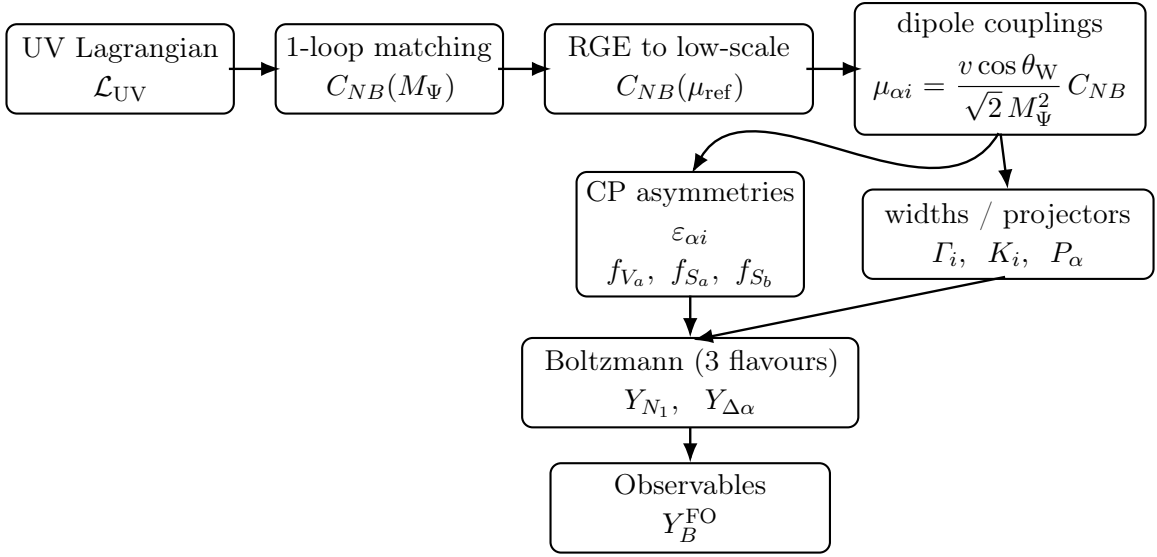
\begin{figure}[t]
\centering
\begin{adjustbox}{max width=\linewidth}
\begin{tikzpicture}[node distance=0.035\linewidth and 0.035\linewidth]

\node[box] (uv)    {UV Lagrangian\\$\mathcal{L}_{\rm UV}$};
\node[box, right=of uv] (match) {1-loop matching\\$C_{NB}(M_\Psi)$};
\node[box, right=of match, text width=0.22\linewidth] (rge) {RGE to low-scale\\$C_{NB}(\mu_{\rm ref})$};
\node[box, right=of rge, text width=0.24\linewidth] (dip) {dipole couplings\\[0.5em]
$\displaystyle \mu_{\alpha i}=\frac{v\cos\theta_{\rm W}}{\sqrt{2}\,M_\Psi^{2}}\,C_{NB}$};

\node[box, below=0.05\linewidth of rge, xshift=0.01\linewidth] (cp) {CP asymmetries\\
$\varepsilon_{\alpha i}$\\ $f_{V_a},\; f_{S_a},\; f_{S_b}$};
\node[box, right=0.05\linewidth of cp, text width=0.24\linewidth] (wd) {widths / projectors\\
$\varGamma_i,\;K_i,\;P_\alpha$};

\node[box, below=0.035\linewidth of cp, text width=0.28\linewidth] (be) {Boltzmann (3 flavours)\\
$Y_{N_1},\;Y_{\Delta\alpha}$};
\node[box, below=0.03\linewidth of be,  text width=0.23\linewidth] (obs) {Observables\\$Y_B^{\rm FO}$};

\makeports{uv}\makeports{match}\makeports{rge}\makeports{dip}
\makeports{cp}\makeports{wd}\makeports{be}\makeports{obs}

\draw[arr] (uvE)    -- (matchW);
\draw[arr] (matchE) -- (rgeW);
\draw[arr] (rgeE)   -- (dipW);

\draw[arr] (dipS) -- (wdN);                  
\draw[arr] (dipS) to[out=240,in=60] (cpN);  

\draw[arr] (wdS) -- (beN);   
\draw[arr] (cpS) -- (beN);   
\draw[arr] (beS) -- (obsN);  

\end{tikzpicture}
\end{adjustbox}
\caption{Workflow of the EFT-consistent analysis adopted in this work.}
\label{fig:1}
\end{figure}

In this work, we show the following:
\begin{itemize}
\item An end-to-end EFT formulation. We formulate EMLG within EFT and perform an end-to-end calculation (see also fig.~\ref{fig:1}).

\item In our setup, the only contributing Wilson coefficient is $\CNB$. We find no one-loop mixing of $\ONB$ with $\mathcal{O}_{NW}$ (i.e. $Z_{BW}=1$), so the analysis cleanly isolates the $\mathrm{U}(1)_Y$ dipole operator.


\item Origin of the suppression. Gauge invariance forces a Higgs insertion in the dimension-six operator, which yields the broken-phase dipole couplings to scale as $\mu\propto v/M_{\Psi}^2$ for fixed heavy-mass ratios $\{r,\xi_j\}$, so that both $\varGamma$ and $\varepsilon$ inherit a common $v^2/M_{\Psi}^4$ suppression.
\end{itemize}

This paper is organised as follows: In sec.~\ref{sec:2}, we obtain the dipole operator $\ONB$ and the Wilson coefficient $\CNB$ by one-loop matching at a high scale. In sec.~\ref{sec:3} we derive the RGE for $\CNB$. Sec.~\ref{sec:4} works in the broken phase and matches $\ONB$ onto the dipole couplings. We compute two-body widths and CP asymmetries. In sec.~\ref{sec:5}, we examine the dynamics of EMLG, flavour effects, and sphaleron transitions. Sec.~\ref{sec:6} presents the fully flavoured Boltzmann system and its solutions, and compares the freeze-out baryon asymmetry with the observed value. In sec.~\ref{sec:7} we highlight the suppression and its underlying origin, and examine the high-scale EMLG.
We also consider possible strategies required for low-scale EMLG to reproduce the observed asymmetry. Finally, in sec.~\ref{sec:8} we briefly summarise the results and outline prospects for future work.

\section{UV--EFT one-loop matching and Wilson coefficients}
\label{sec:2}

In this section we start from a renormalisable (dimension-four) UV Lagrangian with heavy fields and integrate them out to obtain gauge-invariant dimension-six effective operators together with their Wilson coefficients. 

In what follows we work in the mass eigenbasis of the right-handed neutrinos. The Majorana mass matrix $M_N$ is brought to diagonal form by an Autonne--Takagi decomposition with a unitary matrix $U_N$, $U_N^{\intercal}M_NU_N=\diag(M_1,\ldots,M_{n_N})$, and we denote the mass eigenstates by $N_i$ ($i=1,\ldots,n_N$). Unless otherwise stated, all Yukawa matrices and Wilson coefficients appearing below are written in this basis.

\subsection{UV Lagrangian}
\label{sec:2.1}

Here we introduce a set of new fields, extending the SM---right-handed neutrinos $N_i$ (with mass $M_i$), a vector-like fermion $\Psi$ (with mass $M_{\Psi}$), and a charged scalar $S$ (with mass $M_S$). Let $D_{\mu}$ denote the covariant derivative, and let $\lambda_i$ and $\lambda'_i$ denote the Yukawa couplings for the $N_i$-$\Psi$-$S$ interactions, while $y_{\alpha j}$ is the SM Dirac Yukawa coupling for $L_{\alpha}$-$\tilde{H}$-$N_j$. We assume that $\Psi$ carries no flavour index; accordingly, the collections $(\lambda_i)$ and $(\lambda'_i)$ form vectors in the right-handed-neutrino flavour space. We focus on the following UV Lagrangian~\cite{PhysRevD.80.013010}:
\beq
\label{eq:2.1}
\begin{split}
\mathcal{L}_{\rm UV}=\mathcal{L}_{\rm SM}
&+\bar{N}_i\mathrm{i}\slashed{\partial}N_i
+\bar{\Psi}(\mathrm{i}\slashed{D}-M_{\Psi})\Psi
+\abs{D_{\mu}S}^2-V(H,S)\\[0.25em]
&-\pqty*{\lambda'_i\bar{N}_i\Psi_RS+\lambda_i\bar{\Psi}_LN_iS^{\dagger}
+\dfrac{1}{2}\bar{N}M_NN}\\[0.25em]
&-\pqty*{y_{\alpha j}\bar{L}_{\alpha}\tilde{H}N_j
+y_{j\alpha}^{\ast}\bar{N}_j\tilde{H}^{\dagger}L_{\alpha}},
\end{split}
\eeq
with the scalar potential
\beq
\label{eq:2.2}
V(H,S)=M_S^2\abs{S}^2
+\lambda_S\abs{S}^4+\lambda_{HS}\abs{H}^2\abs{S}^2.
\eeq
As is evident from eqs.~(\ref{eq:2.1})--(\ref{eq:2.2}), the UV model is gauge-invariant and renormalisable. 

\begin{table}[t]
\centering
\begin{tabular}{ccccc}\hline
species&quantum numbers&$\mathrm{SU}(3)_c$&$\mathrm{SU}(2)_L$&$\mathrm{U}(1)_Y$\\\hline
$N_i$&$(\bm{1},\bm{1},0)$&$\bm{1}$&$\bm{1}$&0\\
$\Psi$&$(\bm{1},\bm{1},+1)$&$\bm{1}$&$\bm{1}$&$+1$\\
$S$&$(\bm{1},\bm{1},-1)$&$\bm{1}$&$\bm{1}$&$-1$\\\hline
\end{tabular}
\caption{Quantum numbers of the additional fields. In particular, $\Psi$ and $S$ are $\mathrm{SU}(2)_L$ singlets and therefore do not couple to the $\mathrm{SU}(2)_L$ gauge bosons $W_{\mu}^I$ ($I=1,2,3$); no such vertices appear in fig.~\ref{fig:2}.}
\label{tab:1}
\end{table}

From table~\ref{tab:1} one sees that $\Psi$ is a $\mathrm{SU}(2)_L$ singlet with hypercharge $Y_{\Psi}=+1$ and $S$ is a $\mathrm{SU}(2)_L$ singlet with $Y_S=-1$, so indeed $Y_S=-Y_{\Psi}$. Consequently, the operator $\bar{N}\Psi S$ is gauge invariant, and a one-loop graph with the $\mathrm{U}(1)_Y$ gauge boson generates the effective operator $(\bar{L}\sigma^{\mu\nu}N)\tilde{H}B_{\mu\nu}$. The hypercharges $Y_{\Psi}=+1$ and $Y_S=-1$ forbid terms such as $S^2$ ($Y=-2$) or $\bar{\Psi}\tilde{H}$ ($Y=-3/2$), therefore no additional discrete symmetry (e.g. $Z_2$) is required.

\subsection{Integrating out the heavy fields $\Psi$ and $S$}
\label{sec:2.2}

We match the UV model onto the gauge-invariant dipole operator
\beq
\label{eq:2.3}
\mathcal{O}_{NB,\alpha i}
=(\bar{L}_{\alpha}\sigma^{\mu\nu}P_RN_i)\tilde{H}B_{\mu\nu},\qquad
\sigma^{\mu\nu}\coloneqq\dfrac{\mathrm{i}}{2}[\gamma^{\mu},\gamma^{\nu}],
\qquad
\tilde{H}\coloneqq\mathrm{i}\sigma_2H^{\ast},
\eeq
with $B_{\mu\nu}=\partial_{\mu}B_{\nu}-\partial_{\nu}B_{\mu}$. In our UV completion $\Psi$ and $S$ are $\mathrm{SU}(2)_L$ singlets; hence $\cO_{NW}$ is not generated at one loop. In the symmetric phase, we also find no operator mixing into $\cO_{NW}$ at one loop ($B$--$W^3$ kinetic mixing vanishes for complete $\mathrm{SU}(2)_L$ multiplets; see sec.~\ref{sec:3}), so only $\CNB$ contributes at this order.

We work in dimensional regularisation with the $\overline{\rm MS}$ scheme and treat the external charged-lepton and neutrino masses as negligible. The one-loop vertex function $\varGamma^{\mu}$ can be decomposed as
\beq
\label{eq:2.4}
\varGamma^{\mu}(q)=A(q^2)\gamma^{\mu}+B(q^2)q^{\mu}
+C(q^2)\,\mathrm{i}\sigma^{\mu\nu}q_{\nu}+\cO(q^2),
\eeq
where $q$ denotes the four-momentum of the $\mathrm{U}(1)_Y$ gauge boson $B_{\mu}$. Gauge invariance implies the transversality condition $q_{\mu}\varGamma^{\mu}(q)=0$. In the on-shell limit for the emitted gauge boson, $q^2=0$, and using the equations of motion for the external spinors, the constant $\gamma^{\mu}$ piece is forced to vanish in the sum of graphs; hence the first non-vanishing contribution relevant for matching is the dipole term $C(0)\,\mathrm{i}\sigma^{\mu\nu}q_{\nu}$. Emission from the scalar line is purely longitudinal and cancels by the Ward--Takahashi identity; lacking any $\gamma$-matrix structure, it does not project onto $\sigma^{\mu\nu}$, so the dipole is induced only by emission from the fermion line.


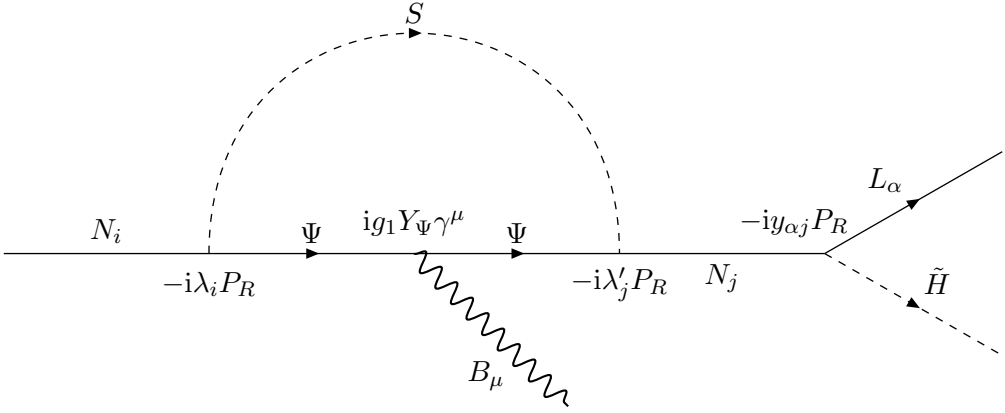
\begin{figure}[t]
\centering
\begin{tikzpicture}[baseline=-0.1cm,scale=1.35]
\begin{feynhand}
\tikzset{
    wavy/.style={decorate, decoration={snake, amplitude=1.25mm, segment length=2.9mm}, thick},
}
\vertex (i1) at (-4,0);
\vertex (w2) at (-2,0);
\vertex (w3) at (0,0);
\vertex (w4) at (2,0);
\vertex (f5) at (4,0);
\vertex (f6) at (1.5,-1.5);
\vertex (f7) at (5.73,-1);
\vertex (f8) at (5.73,1);
\propag [plain] (i1) to [edge label=$N_i$] (w2);
\propag [fer] (w2) to [edge label=$\Psi $] (w3);
\propag [fer] (w3) to [edge label=$\Psi $] (w4);
\draw [wavy] (w3) to (f6);
\propag [chasca] (w2) to [edge label=$S$, half left, looseness=1.85] (w4);
\propag [plain] (w4) to [edge label'=$N_j$] (f5);
\propag [fer] (f5) to [edge label=$L_{\alpha}$] (f8);
\propag [chasca] (f5) to [edge label=$\tilde{H}$] (f7);
\node at (-2,-0.3) {$-\mathrm{i}\lambda_iP_R$};
\node at (2,-0.3) {$-\mathrm{i}\lambda'_jP_R$};
\node at (0,0.3) {$\mathrm{i}g_1Y_{\Psi}\gamma^{\mu}$};
\node at (0.7,-1.2) {$B_{\mu}$};
\node at (3.7,0.3) {$-\mathrm{i}y_{\alpha j}P_R$};
\end{feynhand}
\end{tikzpicture}
\caption{First one-loop matching graph for $\mathcal{O}_{NB}$. The $\mathrm{U}(1)_Y$ gauge boson $B_{\mu}$ is emitted from the $\Psi$ line through $+\mathrm{i}g_1Y_{\Psi}\gamma^{\mu}$. The loop contains $\Psi$ and $S$ and attaches via $-\mathrm{i}\lambda_iP_R$ and $-\mathrm{i}\lambda'_jP_R$. The external process is $N_i\to L_{\alpha}+\tilde{H}+B_{\mu}$. The mirror graph $i\leftrightarrow j$ produces the antisymmetric combination $(\lambda_i\lambda'_j-\lambda_j\lambda'_i)$. The amplitude is UV-finite; emission from the scalar line is purely longitudinal and cancels by the Ward--Takahashi identity, so only the fermion line induces the dipole.}
\label{fig:2}
\end{figure}

At the matching scale $\mu=M_{\Psi}$, the one-loop Wilson coefficient is obtained from the pair of mirror graphs with $\Psi$ and $S$ running in the loop~\cite{PhysRevD.80.013010, Choudhury:2011gbi, Patra:2011aa}. The relevant $\Psi$-$\Psi$-$B$ vertex $+\mathrm{i}g_1Y_{\Psi}\gamma^{\mu}$ ($g_1\coloneqq g'$ denotes the $\mathrm{U}(1)_Y$ gauge coupling) originates from
\beq
\label{eq:2.5}
\mathcal{L}_{\rm UV}\supset g_1Y_{\Psi}\bar{\Psi}\gamma^{\mu}\Psi B_{\mu},
\eeq
while the $S$-$S$-$B$ vertex is $+\mathrm{i}g_1Y_S(2k+q)^{\mu}$. By the Gordon decomposition, the longitudinal term cancels against the Ward--Takahashi identity, leaving only the dipole structure. At the matching scale $\mu=M_{\Psi}$, the sum of the two mirror graphs gives the on-shell amplitude (see app.~\ref{sec:A} for details)
\begin{align}
\mathcal{M}_{\rm full}
&=\dfrac{C_{NB,\alpha i}^{\rm full}(M_{\Psi})}{M_{\Psi}^2}\,
\bar{u}_{L_{\alpha}}\sigma^{\mu\nu}q_{\nu}P_Ru_{N_i}\varepsilon_{\mu}.
\label{eq:2.6}
\end{align}
Since the full amplitude $\mathcal{M}_{\rm full}$ carries the Lorentz structure $\bar{L}\sigma^{\mu\nu}P_RN\times q_{\nu}\varepsilon_{\mu}$, inserting the Higgs doublet $\tilde{H}$ to restore gauge invariance yields the effective operator
\beq
\label{eq:2.7}
\mathcal{O}_{NB,\alpha i}
=(\bar{L}_{\alpha}\sigma^{\mu\nu}P_RN_i)\tilde{H}B_{\mu\nu}.
\eeq
This completes the derivation of the dipole operator obtained by integrating out the heavy fields.

\subsection{One-loop matching for $\ONB$ and $\CNB$}
\label{sec:2.3}

We now perform the one-loop matching~\cite{Schwartz_2013} by computing the full-theory amplitude with external legs $N_i(k)$, $L_{\alpha}(p)$, $B_{\mu}(q)$, and $\tilde{H}(q_H=0)$, and comparing it with the amplitude generated by the EFT operator with the same external legs.
We extract the EFT Wilson coefficient by comparing the EFT amplitude generated by $\ONB$ with the full one-loop amplitude for those external states. On the EFT side, the tree-level amplitude induced by $\ONB$ for the external legs takes the form~(cf. app.~\ref{sec:A})
\beq
\label{eq:2.8}
\mathcal{M}_{\rm EFT}
=\dfrac{C_{NB,\alpha i}^{\rm EFT}(M_{\Psi})}{M_{\Psi}^2}\,
\bar{u}_{L_{\alpha}}(p)\sigma^{\mu\nu}q_{\nu}P_Ru_{N_i}(k)
\varepsilon_{\mu}(q)\tilde{H}(q_H=0).
\eeq
After integrating out the heavy fields, the low-energy $S$-matrix must reproduce the full theory. Hence, the one-loop full-theory amplitude matches the EFT tree amplitude, yielding
\beq
\label{eq:2.9}
C_{NB,\alpha i}^{\rm EFT}(M_{\Psi})
=C_{NB,\alpha i}^{\rm full}(M_{\Psi}),
\eeq
and therefore
\beq
\label{eq:2.10}
C_{NB,\alpha i}^{\rm EFT}(M_{\Psi})
=\sum_{j}
\dfrac{g_1Y_{\Psi}(\lambda_i\lambda'_j-\lambda_j\lambda'_i)y_{\alpha j}}{16\pi^2}
\cdot\dfrac{M_{\Psi}}{M_j}
\cdot\dfrac{1-r+r\ln r}{(1-r)^2},
\eeq
where $r\coloneqq M_S^2/M_{\Psi}^2$ and $\xi_j\coloneqq M_{\Psi}/M_j$ are dimensionless heavy-mass ratios. The flavour structure of $\CNB$ is antisymmetric in the right-handed-neutrino indices, $C_{NB,\alpha i}\propto (\lambda_i\lambda'_j-\lambda_j\lambda'_i)y_{\alpha j}$, so only off-diagonal pairs with $i\neq j$ contribute. In particular, for a single right-handed-neutrino generation the coefficient vanishes identically; hence, at least two generations of right-handed neutrinos are required for this mechanism to operate.

\section{RG flow in the EFT}
\label{sec:3}

We evolve the Wilson coefficient of the dipole operator from the matching scale $\mu=M_{\Psi}$ down to an electroweak reference scale $\mu_{\rm ref}$. Throughout, we work in dimensional regularisation with $n=4-2\epsilon$, employ the $\overline{\rm MS}$ scheme, and use the 't Hooft--Feynman gauge $\xi=1$; see app.~\ref{sec:B}. Light external fermion masses are neglected. All running couplings are evolved at one loop unless otherwise stated.

\subsection{Executive summary and setup}
\label{sec:3.1}

At the matching scale, we obtained~(\ref{eq:2.7}) and (\ref{eq:2.10}).
Renormalisation in the EFT is implemented as follows. First, the external fields are renormalised as
\beq
\label{eq:3.1}
L_0=Z_L^{1/2}L,\quad
N_0=Z_N^{1/2}N,\quad
H_0=Z_H^{1/2}H,\quad
B_{0\mu}=Z_B^{1/2}B_{\mu}.
\eeq
Second, the composite dipole operator is renormalised in the all-legs scheme: the renormalisation constants from the external legs are absorbed into the Wilson coefficient,
\beq
\label{eq:3.2}
C_{0NB}=Z_{NB}C_{NB},\qquad
Z_{NB}\coloneqq (Z_LZ_NZ_HZ_B)^{-1/2},
\eeq
while the bare operator satisfies $\mathcal{O}_{0NB}
=(Z_LZ_NZ_HZ_B)^{1/2}\ONB$. Hence, the bare product 
\beq
\label{eq:3.3}
C_{0NB}\mathcal{O}_{0NB}=\CNB\ONB
\eeq
is finite at one loop, and no additional counterterm for the operator itself is required. The one-loop UV poles relevant to $C_{NB}$ arise from the graphs listed in table~\ref{tab:2}. A representative derivation is presented in sec.~\ref{sec:3.2}, where the off-diagonal counterterm is determined. The assembly of the anomalous dimension then follows in sec.~\ref{sec:3.3}.

\subsection{Off-diagonal gauge mixing}
\label{sec:3.2}

\begin{table}[t]
\centering
\begin{tabular}{llcr}\hline
label&loop&external legs
&UV pole of $\delta Z^{(1)}$ in units of $1/(16\pi^2\epsilon)$\\[0.25em]\hline
(a)&$f,H,c$&$B_{\mu}W_{\nu}^3$&0\\[0.25em]
(b)&$f,H$&$B_{\mu}B_{\nu}$&$-\frac{41}{6}g_1^2$\\[0.25em]
(c)&$f,H,W^{\pm},c$&$W_{\mu}^3W_{\nu}^3$&$-\frac{5}{6}g_2^2$\\[0.25em]
(d)&$L\tilde{H}$&$\bar{N}_iN_i$&$-\frac{1}{2}(yy^{\dagger})_{\beta\alpha}$
\\[0.25em]
(e)&$BL,W^IL,N_i\tilde{H}$&$\bar{L}_{\beta}L_{\alpha}$
&$-\frac{1}{4}g_1^2-\frac{3}{4}g_2^2-\frac{1}{2}(yy^{\dagger})_{\beta\alpha}$
\\[0.25em]
(f)&$B\tilde{H},W^I\tilde{H},N_iL,t_RQ_3$&$\tilde{H}^{\dagger}\tilde{H}$
&$-\frac{1}{2}g_1^2-\frac{3}{2}g_2^2+(yy^{\dagger})_{\beta\alpha}+3y_t^2$
\\[0.25em]\hline
\end{tabular}
\caption{One-loop graphs that can generate UV poles in counterterms relevant for the renormalisation of $\ONB$. Here $f$ denotes Dirac fermions, $H$ the Higgs doublet, $c$ the Faddeev--Popov ghost, and $t_RQ_3$ the top-Yukawa insertion. $\delta Z^{(1)}$ denotes the one-loop pole part of the relevant wave-function or mixing counterterm in the $\overline{\rm MS}$ scheme. Had $B$--$W^3$ mixing been non-zero, the graphs in row (c) would have been required. 
}
\label{tab:2}
\end{table}

The UV-divergent one-loop graphs relevant to the renormalisation of the dipole operator are listed in table~\ref{tab:2}. The light-field quantum numbers $Y$ and $T^3$ used in one-loop two-point functions and field renormalisations are summarised in table~\ref{tab:3}. For the off-diagonal two-point function between the $B$ boson and the $W^3$ boson, the UV pole is proportional to the transverse tensor $(g^{\mu\nu}q^2-q^{\mu}q^{\nu})$ times $\sum_{\rm light}YT^3$, so that evaluating $\langle B_{\mu}W_{\nu}^3\rangle$ with all light SM fields (charges listed in table~\ref{tab:3}) simply yields this product. Since
\beq
\label{eq:3.4}
\sum_{\rm light}YT^3=0,
\eeq
the one-loop counterterm vanishes,
\beq
\label{eq:3.5}
Z_{BW}=1,
\eeq
and thus there is no one-loop mixing between $\ONB$ and $\cO_{NW}$.

\subsection{One-loop renormalisation of $\ONB$ and $\CNB$}
\label{sec:3.3}

Using the field renormalisations in~(\ref{eq:3.1}), the Wilson coefficient obeys
\beq
\label{eq:3.6}
\mu\dfrac{\dd}{\dd\mu}C_{NB}
=-2\gamma_{NB}C_{NB},\qquad
\gamma_{NB}
=\dfrac{1}{2}(\gamma_L+\gamma_N+\gamma_H+\gamma_B),
\eeq
where $\gamma_X=\frac{1}{2}\mu\frac{\partial}{\partial\mu}\ln Z_X|_{g_0,\epsilon}$~\cite{Kugo1989II, Schwartz_2013} are the field anomalous dimensions. From the one-loop two-point functions listed in table~\ref{tab:2} we read off
\beq
\label{eq:3.7}
\begin{split}
\delta Z_L^{(1)}
&=\dfrac{1}{16\pi^2}\epsilon^{-1}
\bqty*{\pqty*{-\dfrac{1}{4}g_1^2-\dfrac{3}{4}g_2^2}\bm{1}
-\dfrac{1}{2}yy^{\dagger}},\quad
\delta Z_N^{(1)}
=\dfrac{1}{16\pi^2}\epsilon^{-1}\cdot\pqty*{-\dfrac{1}{2}yy^{\dagger}},
\\[0.5em]
\delta Z_H^{(1)}
&=\dfrac{1}{16\pi^2}\epsilon^{-1}
\bqty*{\pqty*{-\dfrac{1}{2}g_1^2-\dfrac{3}{2}g_2^2+3y_t^2}\bm{1}
+yy^{\dagger}},\quad
\delta Z_B^{(1)}
=\dfrac{1}{16\pi^2}\epsilon^{-1}\cdot\pqty*{-\dfrac{41}{6}g_1^2}\bm{1},
\end{split}
\eeq
where $\bm{1}$ acts in right-handed-neutrino flavour space. Using~(\ref{eq:3.6}) and~(\ref{eq:3.7}), the one-loop anomalous dimension is
\beq
\label{eq:3.8}
\gamma_{NB}^{(1)}
=\dfrac{1}{32\pi^2}
\pqty*{\dfrac{91}{12}g_1^2+\dfrac{9}{4}g_2^2-3y_t^2}\bm{1},
\eeq
and the Wilson coefficient satisfies the RGE
\beq
\label{eq:3.9}
\mu\dfrac{\dd}{\dd\mu}C_{NB,\alpha i}(\mu)
=-\dfrac{1}{16\pi^2}
\pqty*{\dfrac{91}{12}g_1^2+\dfrac{9}{4}g_2^2-3y_t^2}C_{NB,\alpha i}(\mu).
\eeq
Formally,
\beq
\label{eq:3.10}
\CNB(\mu_{\rm ref})=\CNB(M_{\Psi})
\exp\bqty*{-\dfrac{1}{16\pi^2}\int_{\ln M_{\Psi}}^{\ln\mu_{\rm ref}}
A(\mu)\,\dd\ln\mu},
\eeq
where $A(\mu)$ is defined as
\beq
\label{eq:3.11}
A(\mu)\coloneqq
\dfrac{91}{12}g_1^2(\mu)+\dfrac{9}{4}g_2^2(\mu)-3y_t^2(\mu).
\eeq
The resulting RGE~(\ref{eq:3.9}) is gauge-parameter-independent and consistent with the absence of one-loop $B$--$W^3$ mixing (i.e. $Z_{BW}=1$).
Evolving the Wilson coefficient $\CNB$ from $M_{\Psi}\sim 10~{\rm TeV}$ down to $\mu_{\rm ref}=150~{\rm GeV}$ with the one-loop RGE modifies $\abs{\CNB}$ typically at the $\cO(10\%)$ level.\footnote{See also refs.~\cite{Jenkins:2013wua, Buchalla:2019wsc} for general SMEFT operator-mixing structures.}

\begin{table}[t]
\centering
\begin{tabular}{lrrrr}\hline
light field&multiplicity&$Y$&$T^3$&$YT^3$\\[0.1em]\hline
$(u,d)_L$&$2\times 3$&$+\frac{1}{6}$&$\pm\frac{1}{2}$&0\\[0.1em]
$u_R$&$1\times 3$&$+\frac{2}{3}$&0&0\\[0.1em]
$d_R$&$1\times 3$&$-\frac{1}{3}$&0&0\\[0.1em]
$(\nu,e)_L$&2&$-\frac{1}{2}$&$\pm\frac{1}{2}$&0\\[0.1em]
$e_R$&1&$-1$&0&0\\[0.1em]
$H$&2&$+\frac{1}{2}$&$\pm\frac{1}{2}$&0\\[0.1em]\hline
\end{tabular}
\caption{Light-field charges used in the calculations of two-point functions in table~\ref{tab:2}. Entries are given per Weyl fermion.}
\label{tab:3}
\end{table}

\subsection{Coupled running of gauge and Yukawa couplings}
\label{sec:3.4}

The one-loop RGE for the Wilson coefficient depends on $A(\mu)$ in eq.~(\ref{eq:3.11}) through
\beq
\label{eq:3.12}
\mu\dfrac{\dd}{\dd\mu}\CNB(\mu)
=-\dfrac{1}{16\pi^2}A(\mu)\CNB(\mu).
\eeq
Hence, the evolution of $\CNB$ is quantitatively controlled by the accuracy with which $y_t(\mu)$ is evolved. In particular, the one-loop beta function for the top Yukawa coupling contains a comparatively large gluonic term~\cite{MACHACEK1984221},\footnote{The original work adopted the $\mathrm{SU}(5)$ normalisation for the hypercharge coupling, $g_1=\sqrt{3/5}\,g_1'$~\cite{MACHACEK198383}. In this paper we use the canonical normalisation. Accordingly, the numerical factor quoted in~\cite{MACHACEK1984221} is rewritten as $\frac{17}{20}g_1'^2=\frac{17}{20}\cdot\frac{5}{3}g_1^2=\frac{17}{12}g_1^2$ (see eq.~(\ref{eq:3.13})).}
\beq
\label{eq:3.13}
\mu\dv{y_t}{\mu}
=\dfrac{y_t}{16\pi^2}
\bqty*{\dfrac{9}{2}y_t^2
-\pqty*{\dfrac{17}{12}g_1^2
+\dfrac{9}{4}g_2^2+8g_3^2}},
\eeq
thus freezing $g_3$ and solving the RGE would bias $y_t^3(\mu)$. Numerically, for typical boundary conditions we find
\beq
\label{eq:3.14}
g_3(M_{\Psi}=10~{\rm TeV})\simeq 0.957,\qquad
g_3(\mu_{\rm ref}=150~{\rm GeV})\simeq 1.17,
\eeq
i.e. a sizeable change along the flow. If $g_3$ were kept fixed, the solution for $y_t^2(\mu_{\rm ref})$ would be underestimated at $\cO(10\%)$ level. Since $A(\mu)$ contains the term $-3y_t^2$, an underestimate of $y_t^2$ induces a systematic shift of $\mathcal{O}(10\%)$ in $A(\mu)$; this shift then accumulates in the exponential running of $\CNB$. Furthermore, $A(\mu)$ contains the gauge couplings $g_1$ and $g_2$. For this reason, all our numerical results jointly evolve $\{g_1,g_2,g_3,y_t\}$ together with $\CNB$ between $\mu=M_{\Psi}$ and $\mu_{\rm ref}=150~{\rm GeV}$.


\section{Decay widths and CP asymmetries}
\label{sec:4}

In this section we compute the flavour-dependent decay widths and CP asymmetries. Ref.~\cite{PhysRevD.78.085024} analysed the two-body decay $N\to\nu+\gamma$ induced by a dimension-five dipole operator but did not include the $Z$ mode $N\to\nu+Z$. Since the latter carries non-negligible phase-space and mixing-angle effects, we include both channels in order to obtain a quantitatively robust assessment.

\subsection{Dipole coupling and decay modes}
\label{sec:4.1}

After EWSB, the gauge-invariant dimension-six operator $\ONB$ maps onto a dimension-five dipole operator. Using $\tilde{H}\to(v/\sqrt{2}\;\;0)^{\intercal}$ and expressing $B_{\mu\nu}$ in the physical basis, $B_{\mu\nu}=F_{\mu\nu}\cos\theta_{\rm W}-Z_{\mu\nu}\sin\theta_{\rm W}$, the effective interaction takes the form
\beq
\label{eq:4.1}
-\mathcal{L}_{\rm EFT}\supset
C_{NB,\alpha i}(\mu_{\rm ref})\dfrac{v}{\sqrt{2}\,M_{\Psi}^2}
(\bar{\nu}_{\alpha}\sigma^{\mu\nu}P_RN_i)
(F_{\mu\nu}\cos\theta_{\rm W}-Z_{\mu\nu}\sin\theta_{\rm W})
+\text{h.c.}
\eeq
(see, e.g.,~\cite{Grzadkowski:2010es, BRIVIO20191}.)
The Wilson coefficient has already been evolved from $M_{\Psi}$ down to $\mu_{\rm ref}$ in sec.~\ref{sec:3}. Throughout this work, we take the electroweak reference scale to be
\beq
\label{eq:4.2}
\mu_{\rm ref}=150~{\rm GeV}.
\eeq
The gauge couplings at $\mu_{\rm ref}=150~{\rm GeV}$ differ from their values at $m_Z\simeq 91.2~{\rm GeV}$ by less than 1\%, and this difference is within the accuracy of our one-loop analysis. Hence, we take $\mu_{\rm ref}=150~{\rm GeV}$ as the reference scale for our numerics. 

It is convenient to define the dipole coupling 
\begin{align}
-\mathcal{L}_{\rm EFT}&\supset
\mu_{\alpha i}(\bar{\nu}_{\alpha}\sigma^{\mu\nu}P_RN_i)
\pqty*{F_{\mu\nu}-Z_{\mu\nu}\tan\theta_{\rm W}}
+\text{h.c.},
\label{eq:4.3}\\[0.25em]
\mu_{\alpha i}&\coloneqq
\dfrac{v\cos\theta_{\rm W}}{\sqrt{2}\,M_{\Psi}^2}\,
C_{NB,\alpha i}(\mu_{\rm ref}).
\label{eq:4.4}
\end{align}
With this normalisation, $\mu_{\alpha i}$ has mass dimension $-1$, as in classical electromagnetism.

Taking the ratio of partial widths yields
\beq
\label{eq:4.5}
\dfrac{\varGamma_Z^{\rm tree}}{\varGamma_{\gamma}^{\rm tree}}
=(1-r_Z)^2(1+r_Z)\tan^2\theta_{\rm W},\qquad
r_Z\coloneqq\dfrac{m_Z^2}{M_i^2}.
\eeq
The right-hand side factorises into a phase-space part and the mixing-angle factor. In the limit $M_i\gg m_Z$, $\varGamma_Z^{\rm tree}/\varGamma_{\gamma}^{\rm tree}\to\tan^2\theta_{\rm W}$. For our benchmark $M_i=500\,{\rm GeV}$ we obtain the branching ratio $B(N\to\nu Z)\simeq 29\%$, i.e. the $Z$ mode is subleading but non-negligible and will be included henceforth.

\subsection{Decay widths}
\label{sec:4.2}

\begin{figure}[t]
\centering
\begin{tabular}{cc}\\
\centering
\begin{tikzpicture}[baseline=-0.1cm,scale=0.8]
\begin{feynhand}
\tikzset{
    wavy/.style={decorate, decoration={snake, amplitude=0.75mm, segment length=2.9mm}, thick},
}
\setlength{\feynhandblobsize}{3mm}
\vertex (i1) at (-3,0);
\vertex [grayblob] (w2) at (0,0) {};
\vertex (f3) at (3,1.5);
\vertex (f4) at (3,-1.5);
\propag [plain] (i1) to [edge label=$N_i(k)$] (w2);
\propag [fer] (w2) to [edge label=${\nu}_{\alpha}(p)$] (f3);
\draw [wavy] (w2) to [edge label'=$\gamma_{\rho}(q)$] (f4);
\end{feynhand}
\end{tikzpicture}
\hspace{1em}
&
\hspace{1em}
\begin{tikzpicture}[baseline=-0.1cm,scale=0.8]
\begin{feynhand}
\tikzset{
    wavy/.style={decorate, decoration={snake, amplitude=0.75mm, segment length=2.9mm}, thick},
}
\setlength{\feynhandblobsize}{3mm}
\vertex (i1) at (-3,0);
\vertex [grayblob] (w2) at (0,0) {};
\vertex (f3) at (3,1.5);
\vertex (f4) at (3,-1.5);
\propag [plain] (i1) to [edge label=$N_i(k)$] (w2);
\propag [fer] (w2) to [edge label=${\nu}_{\alpha}(p)$] (f3);
\draw [wavy] (w2) to [edge label'=$Z_{\rho}(q)$] (f4);
\end{feynhand}
\end{tikzpicture}
\\[6em]
\begin{tikzpicture}[baseline=-0.1cm,scale=0.8]
\begin{feynhand}
\tikzset{
    wavy/.style={decorate, decoration={snake, amplitude=0.75mm, segment length=2.9mm}, thick},
}
\setlength{\feynhandblobsize}{3mm}
\vertex (i1) at (-3,0);
\vertex [grayblob] (w2) at (0,0) {};
\vertex (f3) at (3,1.5);
\vertex (f4) at (3,-1.5);
\propag [plain] (i1) to [edge label=$N_i(k)$] (w2);
\propag [antfer] (w2) to [edge label=$\bar{\nu}_{\alpha}(p)$] (f3);
\draw [wavy] (w2) to [edge label'=$\gamma_{\rho}(q)$] (f4);
\end{feynhand}
\end{tikzpicture}
\hspace{1em}
&
\hspace{1em}
\begin{tikzpicture}[baseline=-0.1cm,scale=0.8]
\begin{feynhand}
\tikzset{
    wavy/.style={decorate, decoration={snake, amplitude=0.75mm, segment length=2.9mm}, thick},
}
\setlength{\feynhandblobsize}{3mm}
\vertex (i1) at (-3,0);
\vertex [grayblob] (w2) at (0,0) {};
\vertex (f3) at (3,1.5);
\vertex (f4) at (3,-1.5);
\propag [plain] (i1) to [edge label=$N_i(k)$] (w2);
\propag [antfer] (w2) to [edge label=$\bar{\nu}_{\alpha}(p)$] (f3);
\draw [wavy] (w2) to [edge label'=$Z_{\rho}(q)$] (f4);
\end{feynhand}
\end{tikzpicture}
\end{tabular}
\vspace{0.5em}
\caption{Two-body decays of a right-handed neutrino $N_i$ induced by the dipole operator $\ONB$, which after EWSB reduces to a dimension-five operator. The wavy line denotes the neutral gauge boson, the photon (left) or the $Z$ boson (right). Decays into final states containing a charged lepton are forbidden by charge conservation. Grey blobs indicate that the effective vertices encode the heavy-field information through the Wilson coefficient $\CNB$.}
\label{fig:3}
\end{figure}
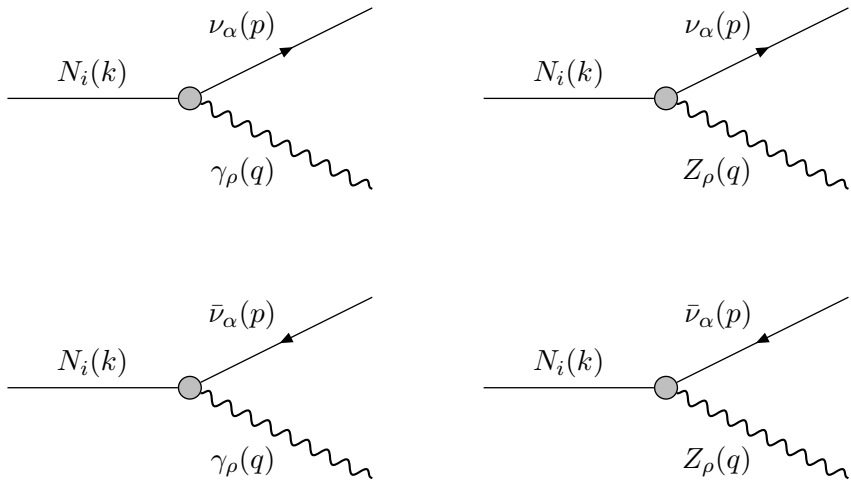

The partial decay width into the photon mode for a given flavour $\alpha=e,\mu,\tau$ reads~\cite{PhysRevD.78.085024, Law:2008yyq, PhysRevD.98.115015}
\beq
\label{eq:4.6}
\varGamma_{\gamma,\alpha i}^{\rm tree}
=\dfrac{\abs{\mu_{\alpha i}}^2M_i^3}{2\pi}
=\dfrac{v^2\cos^2\theta_{\rm W}}{4\pi}
\dfrac{M_i^3}{M_{\Psi}^4}
\abs{C_{NB,\alpha i}(\mu_{\rm ref})}^2,
\eeq
and summing over $\alpha$ gives the total photon-mode width $\varGamma_{\gamma,i}^{\rm tree}$. Using eqs.~(\ref{eq:4.5})--(\ref{eq:4.6}), the partial width into the $Z$ mode per flavour is
\beq
\label{eq:4.7}
\begin{split}
\varGamma_{Z,\alpha i}^{\rm tree}
&=\dfrac{\abs{\mu_{\alpha i}}^2M_i^3}{2\pi}
(1-r_Z)^2(1+r_Z)\tan^2\theta_{\rm W}\\
&=\dfrac{v^2\sin^2\theta_{\rm W}}{4\pi}\dfrac{M_i^3}{M_{\Psi}^4}
(1-r_Z)^2(1+r_Z)
\abs{C_{NB,\alpha i}(\mu_{\rm ref})}^2,
\end{split}
\eeq
and $\varGamma_{Z,i}^{\rm tree}=\sum_{\alpha}\varGamma_{Z,\alpha i}^{\rm tree}$. The total two-body decay width is
\beq
\label{eq:4.8}
\varGamma_i\coloneqq
\varGamma_{\gamma,i}^{\rm tree}+\varGamma_{Z,i}^{\rm tree}.
\eeq
We shall normalise the CP asymmetries by $\varGamma_i$ below.

\subsection{CP asymmetries I: vertex contributions}
\label{sec:4.3}

The CP asymmetry originates from the interference\footnote{The charge conjugate, c.c., in the interference always acts as a projector onto the real part; we take the imaginary part only after forming the CP difference, which is implemented via the discontinuity (optical theorem) in app.~\ref{sec:C.3.1}.}
of the tree amplitude with one-loop corrections built out of the same gauge-invariant operator $\ONB$. Two gauge-parameter-independent topologies contribute: vertex graphs and self-energy (wave-function) graphs.


\begin{figure}[t]
\centering
\begin{minipage}[b]{0.49\hsize}
\centering
\begin{tikzpicture}[baseline=-0.1cm,scale=1]
\begin{feynhand}
\tikzset{
    wavy/.style={decorate, decoration={snake, amplitude=0.75mm, segment length=2.9mm}, thick},
}
\setlength{\feynhandblobsize}{3mm}
\vertex (i1) at (-1.6,0);
\vertex [grayblob] (w2) at (0.4,0) {};
\vertex [grayblob] (w3) at (3,1.5) {};
\vertex [grayblob] (w4) at (3,-1.5) {};
\vertex (f5) at (5,1.5);
\vertex (f6) at (5,-1.5);
\propag [plain] (i1) to [edge label'=$N_i(k)$] (w2);
\propag [antfer] (w2) to [edge label'=$\nu_{\beta}(q_1)$] (w4);
\propag [plain] (w3) to [edge label=$N_m(q_3)$] (w4);
\propag [fer] (w3) to [edge label=$\nu_{\alpha}(p)$] (f5);
\draw [wavy] (w4) to [edge label=$\gamma_{\sigma}(q)$] (f6);
\draw [wavy] (w2) to (w3);
\node at (1.5,1.3) {$\gamma_{\rho}(q_2)$};
\node at (0.2,0.5) {$\tilde{V}^{\rho}_{\beta i}$};
\node at (3,2) {$V_{\alpha m}^{\mu}$};
\node at (3,-2) {$V_{\beta m}^{\sigma}$};
\node at (-1.6,2.5) {\bfseries (a)};
\end{feynhand}
\end{tikzpicture}
\end{minipage}
\begin{minipage}[b]{0.49\hsize}
\centering
\begin{tikzpicture}[baseline=-0.1cm,scale=1]
\begin{feynhand}
\tikzset{
    wavy/.style={decorate, decoration={snake, amplitude=0.75mm, segment length=2.9mm}, thick},
}
\setlength{\feynhandblobsize}{3mm}
\vertex (i1) at (-1.6,0);
\vertex [grayblob] (w2) at (0.4,0) {};
\vertex [grayblob] (w3) at (3,1.5) {};
\vertex [grayblob] (w4) at (3,-1.5) {};
\vertex (f5) at (5,1.5);
\vertex (f6) at (5,-1.5);
\propag [plain, mom'={$k$}] (i1) to (w2);
\propag [antfer, mom'={$q_1$}] (w2) to (w4);
\propag [plain, mom'={$q_3$}] (w4) to (w3);
\propag [fer, mom'={$p$}] (w3) to (f5);
\draw [wavy] (w4) to (f6);
\draw [draw=none, mom={[arrow distance=6pt] {$q$}}] (w4) to (f6);
\draw [wavy] (w2) to (w3);
\draw [draw=none, mom={[arrow distance=6pt] {$q_2$}}] (w2) to (w3);
\node at (1.4,-0.25) {$\beta$};
\node at (1.3,0.2) {$\rho$};
\node at (2.25,-0.75) {$\beta$};
\node at (2.3,0.75) {$\mu$};
\node at (2.7,0.5) {$m$};
\node at (2.7,-0.5) {$m$};
\node at (4, -1.8) {$\sigma$};
\node at (4,1.8) {$\alpha$};
\node at (-0.6,0.3) {$i$};
\end{feynhand}
\end{tikzpicture}
\end{minipage}
\begin{minipage}[b]{0.49\hsize}
\centering
\begin{tikzpicture}[baseline=-0.1cm,scale=1]
\begin{feynhand}
\tikzset{
    wavy/.style={decorate, decoration={snake, amplitude=0.75mm, segment length=2.9mm}, thick},
}
\setlength{\feynhandblobsize}{3mm}
\vertex (i1) at (-1.6,0);
\vertex [grayblob] (w2) at (0.4,0) {};
\vertex [grayblob] (w3) at (3,1.5) {};
\vertex [grayblob] (w4) at (3,-1.5) {};
\vertex (f5) at (5,1.5);
\vertex (f6) at (5,-1.5);
\propag [plain] (i1) to [edge label'=$N_i(k)$] (w2);
\propag [fer] (w2) to [edge label'=$\nu_{\beta}(q_1)$] (w4);
\propag [plain] (w3) to [edge label=$N_m(q_3)$] (w4);
\propag [fer] (w3) to [edge label=$\nu_{\alpha}(p)$] (f5);
\draw [wavy] (w4) to [edge label=$\gamma_{\sigma}(q)$] (f6);
\draw [wavy] (w2) to (w3);
\node at (1.5,1.3) {$\gamma_{\rho}(q_2)$};
\node at (0.25,0.5) {$V^{\rho}_{\beta i}$};
\node at (3,2) {$V_{\alpha m}^{\mu}$};
\node at (3,-2) {$\tilde{V}_{\beta m}^{\sigma}$};
\node at (-1.6,2.5) {\bfseries (b)};
\end{feynhand}
\end{tikzpicture}
\end{minipage}
\begin{minipage}[b]{0.49\hsize}
\centering
\begin{tikzpicture}[baseline=-0.1cm,scale=1]
\begin{feynhand}
\tikzset{
    wavy/.style={decorate, decoration={snake, amplitude=0.75mm, segment length=2.9mm}, thick},
}
\setlength{\feynhandblobsize}{3mm}
\vertex (i1) at (-1.6,0);
\vertex [grayblob] (w2) at (0.4,0) {};
\vertex [grayblob] (w3) at (3,1.5) {};
\vertex [grayblob] (w4) at (3,-1.5) {};
\vertex (f5) at (5,1.5);
\vertex (f6) at (5,-1.5);
\propag [plain, mom'={$k$}] (i1) to (w2);
\propag [fer, mom'={$q_1$}] (w2) to (w4);
\propag [plain, mom'={$q_3$}] (w4) to (w3);
\propag [fer, mom'={$p$}] (w3) to (f5);
\draw [wavy] (w4) to (f6);
\draw [draw=none, mom={[arrow distance=6pt] {$q$}}] (w4) to (f6);
\draw [wavy] (w2) to (w3);
\draw [draw=none, mom={[arrow distance=6pt] {$q_2$}}] (w2) to (w3);
\node at (1.4,-0.25) {$\beta$};
\node at (1.3,0.2) {$\rho$};
\node at (2.25,-0.75) {$\beta$};
\node at (2.3,0.75) {$\mu$};
\node at (2.7,0.5) {$m$};
\node at (2.7,-0.5) {$m$};
\node at (4, -1.8) {$\sigma$};
\node at (4,1.8) {$\alpha$};
\node at (-0.6,0.3) {$i$};
\end{feynhand}
\end{tikzpicture}
\end{minipage}
\caption{Vertex contributions. The contribution is represented by two types of graphs, (a) and (b). Left: particle content, external four-momenta and vertices. Right: the common momentum routing used for the loop integrations.}
\label{fig:4}
\end{figure}
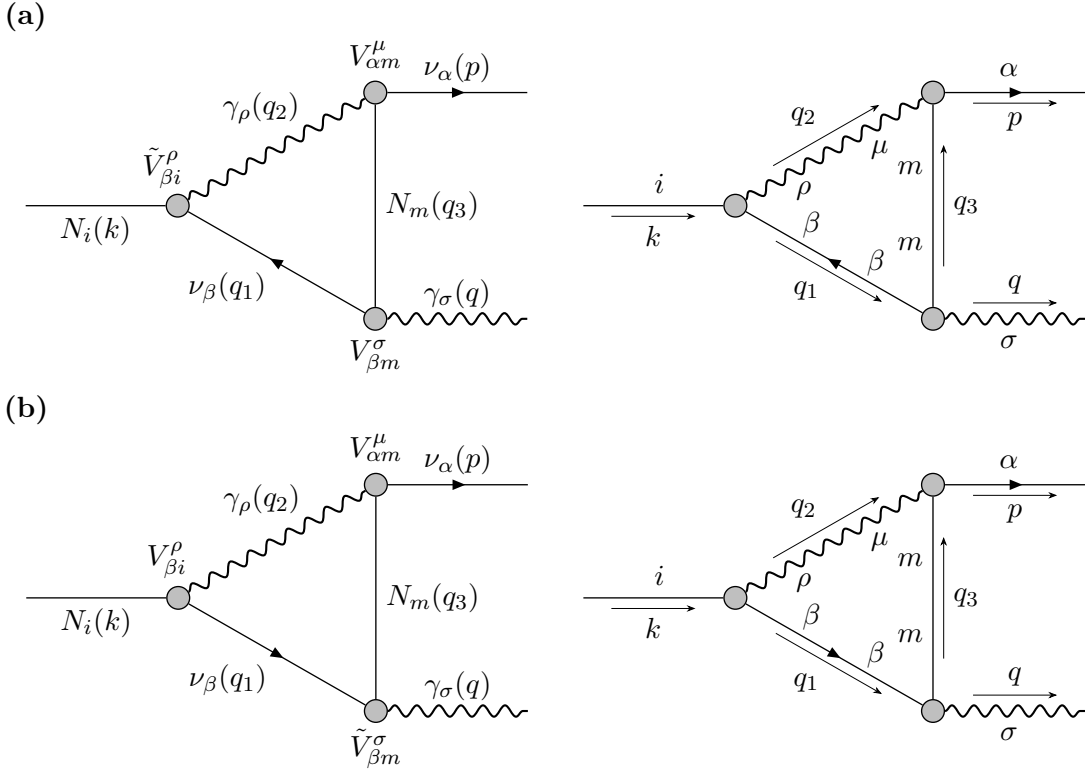

In the centre-of-mass frame, the CP asymmetry associated with the vertex graph in fig.~\ref{fig:4}(a) is
\beq
\label{eq:4.9}
\begin{split}
\varepsilon_{{\rm vert},\alpha i}^{\gamma,\rm (a)}
&=-\dfrac{M_i^2}{2\pi\sum_{\beta}\abs{\mu_{\beta i}}^2}\sum_{m\neq i}
\Im\bqty*{\mu_{\alpha i}^{\ast}\mu_{\alpha m}(\mu^{\dagger}\mu)_{im}}\,
f_{V_a}(x)
\end{split}
\eeq
with $x\coloneqq M_m^2/M_i^2$ and
\beq
\label{eq:4.10}
f_{V_a}(x)\coloneqq
\sqrt{x}\,\Bqty*{1+2x\bqty*{1-(1+x)\ln\dfrac{1+x}{x}}}.
\eeq
For the graph in fig.~\ref{fig:4}(b), the interference term becomes zero.
The total vertex contribution is therefore given by~(\ref{eq:4.9}).

\subsection{CP asymmetries II: self-energy contributions}
\label{sec:4.4}

\begin{figure}[t]
\centering
\begin{minipage}[b]{0.49\hsize}
\centering
\begin{tikzpicture}[baseline=-0.1cm,scale=1]
\begin{feynhand}
\tikzset{
    wavy/.style={decorate, decoration={snake, amplitude=0.75mm, segment length=2.9mm}, thick},
}
\setlength{\feynhandblobsize}{3mm}
\vertex (i1) at (0,0);
\coordinate (w2) at (1.5,0);
\coordinate (w3) at (3.5,0);
\coordinate (w4) at (5.5,0);
\vertex (f5) at (7,1.5);
\vertex (f6) at (7,-1.5);
\propag [plain] (i1) to [edge label=$N_i(k)$] (w2);
\propag [antfer] (w2) to [half right, looseness=1.75, edge label'=$\nu_{\beta}(q_1)$] (w3);
\draw[wavy]
      (w2) arc[start angle=180,end angle=0,radius=1cm]
      node[midway,above] {$\gamma_{\rho}(q_2)$};
\propag [plain] (w3) to [edge label=$N_m(k)$] (w4);
\propag [fer] (w4) to [edge label=$\nu_{\alpha}(p)$] (f5);
\draw [wavy] (w4) to [edge label'=$\gamma_{\sigma}(q)$] (f6);
\node at (1.2,-0.4) {$\tilde{V}_{\beta i}^{\rho}$};
\node at (3.95,-0.4) {$V_{\beta m}^{\mu}$};
\node at (6.25,0) {$V_{\alpha m}^{\sigma}$};
\node at (0,2) {\bfseries (a)};
\vertex [grayblob] (w2) at (1.5,0) {};
\vertex [grayblob] (w3) at (3.5,0) {};
\vertex [grayblob] (w4) at (5.5,0) {};
\end{feynhand}
\end{tikzpicture}
\end{minipage}
\begin{minipage}[b]{0.49\hsize}
\centering
\begin{tikzpicture}[baseline=-0.1cm,scale=1]
\begin{feynhand}
\tikzset{
    wavy/.style={decorate, decoration={snake, amplitude=0.75mm, segment length=2.9mm}, thick},
}
\setlength{\feynhandblobsize}{3mm}
\vertex (i1) at (0,0);
\coordinate (w2) at (1.5,0);
\coordinate (w3) at (3.5,0);
\coordinate (w4) at (5.5,0);
\vertex (f5) at (7,1.5);
\vertex (f6) at (7,-1.5);
\propag [plain, mom'={$k$}] (i1) to [edge label=$i$] (w2);
\propag [antfer, mom'={$q_1$}] (w2) to [half right, looseness=1.75] (w3);
\draw[wavy] 
	(w2) arc[start angle=180,end angle=0,radius=1cm];
\draw[-{Stealth[length=3.25pt,width=2.25pt]},
      line width=0.3pt]
      (2.5,0) ++(135:1.3cm)
      arc[start angle=135, delta angle=-90, radius=1.3cm];
\node at (2.5,1.6) {$q_2$};
\propag [plain, mom'={$k$}] (w3) to (w4);
\propag [fer, mom={$p$}] (w4) to [edge label'=$\alpha$] (f5);
\draw [wavy] (w4) to [edge label=$\sigma$] (f6);
\draw [draw=none, mom'={[arrow distance=6pt] {$q$}}] (w4) to (f6);
\node at (2,-0.55) {$\beta$};
\node at (3,-0.55) {$\beta$};
\node at (2,0.6) {$\rho$};
\node at (3,0.6) {$\mu$};
\node at (4,0.3) {$m$};
\node at (5,0.3) {$m$};
\vertex [grayblob] (w2) at (1.5,0) {};
\vertex [grayblob] (w3) at (3.5,0) {};
\vertex [grayblob] (w4) at (5.5,0) {};
\end{feynhand}
\end{tikzpicture}
\end{minipage}
\begin{minipage}[b]{0.49\hsize}
\centering
\begin{tikzpicture}[baseline=-0.1cm,scale=1]
\begin{feynhand}
\tikzset{
    wavy/.style={decorate, decoration={snake, amplitude=0.75mm, segment length=2.9mm}, thick},
}
\setlength{\feynhandblobsize}{3mm}
\vertex (i1) at (0,0);
\coordinate (w2) at (1.5,0);
\coordinate (w3) at (3.5,0);
\coordinate (w4) at (5.5,0);
\vertex (f5) at (7,1.5);
\vertex (f6) at (7,-1.5);
\propag [plain] (i1) to [edge label=$N_i(k)$] (w2);
\propag [fer] (w2) to [half right, looseness=1.75, edge label'=$\nu_{\beta}(q_1)$] (w3);
\draw[wavy]
      (w2) arc[start angle=180,end angle=0,radius=1cm]
      node[midway,above] {$\gamma_{\rho}(q_2)$};
\propag [plain] (w3) to [edge label=$N_m(k)$] (w4);
\propag [fer] (w4) to [edge label=$\nu_{\alpha}(p)$] (f5);
\draw [wavy] (w4) to [edge label'=$\gamma_{\sigma}(q)$] (f6);
\node at (1.2,-0.4) {$V_{\beta i}^{\rho}$};
\node at (3.95,-0.4) {$\tilde{V}_{\beta m}^{\mu}$};
\node at (6.25,0) {$V_{\alpha m}^{\sigma}$};
\node at (0,2) {\bfseries (b)};
\vertex [grayblob] (w2) at (1.5,0) {};
\vertex [grayblob] (w3) at (3.5,0) {};
\vertex [grayblob] (w4) at (5.5,0) {};
\end{feynhand}
\end{tikzpicture}
\end{minipage}
\begin{minipage}[b]{0.49\hsize}
\centering
\begin{tikzpicture}[baseline=-0.1cm,scale=1]
\begin{feynhand}
\tikzset{
    wavy/.style={decorate, decoration={snake, amplitude=0.75mm, segment length=2.9mm}, thick},
}
\setlength{\feynhandblobsize}{3mm}
\vertex (i1) at (0,0);
\coordinate (w2) at (1.5,0);
\coordinate (w3) at (3.5,0);
\coordinate (w4) at (5.5,0);
\vertex (f5) at (7,1.5);
\vertex (f6) at (7,-1.5);
\propag [plain, mom'={$k$}] (i1) to [edge label=$i$] (w2);
\propag [fer, mom'={$q_1$}] (w2) to [half right, looseness=1.75] (w3);
\draw[wavy] 
	(w2) arc[start angle=180,end angle=0,radius=1cm];
\draw[-{Stealth[length=3.25pt,width=2.25pt]},
      line width=0.3pt]
      (2.5,0) ++(135:1.3cm)
      arc[start angle=135, delta angle=-90, radius=1.3cm];
\node at (2.5,1.6) {$q_2$};
\propag [plain, mom'={$k$}] (w3) to (w4);
\propag [fer, mom={$p$}] (w4) to [edge label'=$\alpha$] (f5);
\draw [wavy] (w4) to [edge label=$\sigma$] (f6);
\draw [draw=none, mom'={[arrow distance=6pt] {$q$}}] (w4) to (f6);
\node at (2,-0.55) {$\beta$};
\node at (3,-0.55) {$\beta$};
\node at (2,0.6) {$\rho$};
\node at (3,0.6) {$\mu$};
\node at (4,0.3) {$m$};
\node at (5,0.3) {$m$};
\vertex [grayblob] (w2) at (1.5,0) {};
\vertex [grayblob] (w3) at (3.5,0) {};
\vertex [grayblob] (w4) at (5.5,0) {};
\end{feynhand}
\end{tikzpicture}
\end{minipage}
\caption{Self-energy contributions. The contribution is likewise given by two types of graphs, (a) and (b). Left: particle content, external four-momenta and vertices. Right: the common momentum routing used for the loop integrations.}
\label{fig:5}
\end{figure}
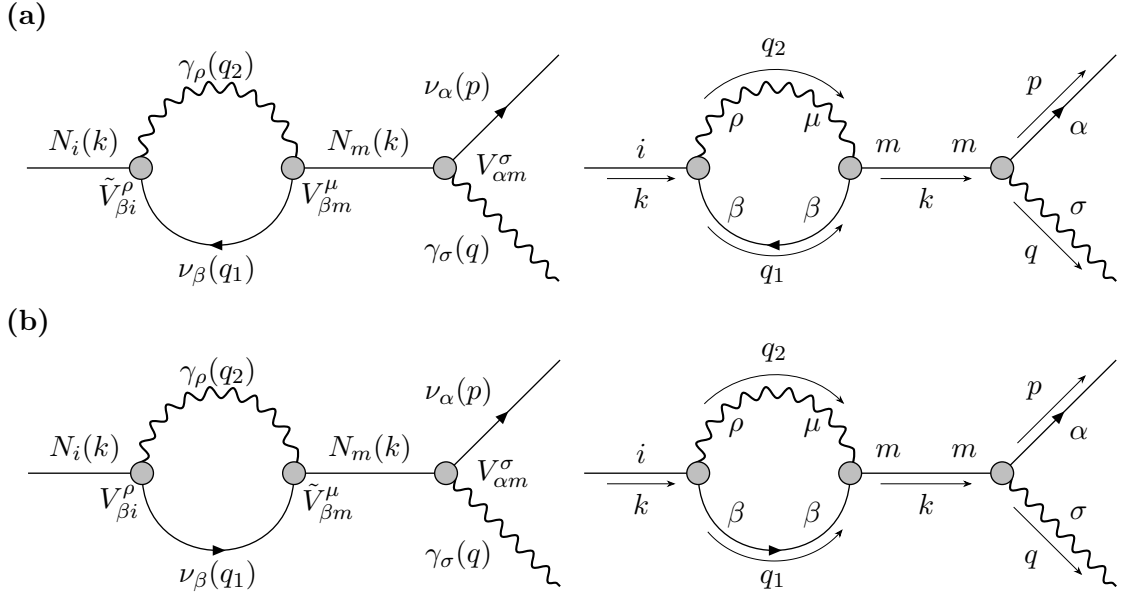

In the centre-of-mass frame, the self-energy contribution associated with fig.~\ref{fig:5}(a) reads
\beq
\label{eq:4.11}
\begin{split}
\varepsilon_{{\rm self},\alpha i}^{\gamma,\rm (a)}
&=-\dfrac{M_i^2}{2\pi\sum_{\beta}\abs{\mu_{\beta i}}^2}\sum_{m\neq i}
\Im\bqty*{\mu_{\alpha i}^{\ast}\mu_{\alpha m}(\mu^{\dagger}\mu)_{im}}\,
\dfrac{\sqrt{x}}{1-x},\qquad
x=\dfrac{M_m^2}{M_i^2},
\end{split}
\eeq
and the contribution of fig.~\ref{fig:5}(b) is
\beq
\label{eq:4.12}
\begin{split}
\varepsilon_{{\rm self},\alpha i}^{\gamma,\rm (b)}
&=-\dfrac{M_i^2}{2\pi\sum_{\beta}\abs{\mu_{\beta i}}^2}\sum_{m\neq i}
\Im\bqty*{\mu_{\alpha i}^{\ast}\mu_{\alpha m}(\mu^{\dagger}\mu)_{mi}}\,
\dfrac{1}{1-x}.
\end{split}
\eeq
Adding the self-energy pieces to the vertex result in eq.~(\ref{eq:4.9}), the photon-mode asymmetry reads
\beq
\label{eq:4.13}
\varepsilon_{\alpha i}^{\gamma}
=-\dfrac{M_i^2}{2\pi\sum_{\beta}\abs{\mu_{\beta i}}^2}\sum_{m\neq i}
\Im\bqty*{\mu_{\alpha i}^{\ast}\mu_{\alpha m}\Bqty*{(\mu^{\dagger}\mu)_{im}
[f_{V_a}(x)+f_{S_a}(x)]+(\mu^{\dagger}\mu)_{mi}f_{S_b}(x)}},
\eeq
with real loop functions
\beq
\label{eq:4.14}
f_{V_a}(x)=\sqrt{x}\,\Bqty*{1+2x\bqty*{1-(1+x)\ln\dfrac{1+x}{x}}},\quad
f_{S_a}(x)=\dfrac{\sqrt{x}}{1-x},\quad
f_{S_b}(x)=\dfrac{1}{1-x}.
\eeq
Here the Majorana back-flow doubles the CP-even interference (cf. app.~\ref{sec:C.3.3}).

\begin{figure}[t]
\centering
\includegraphics[width=\linewidth]{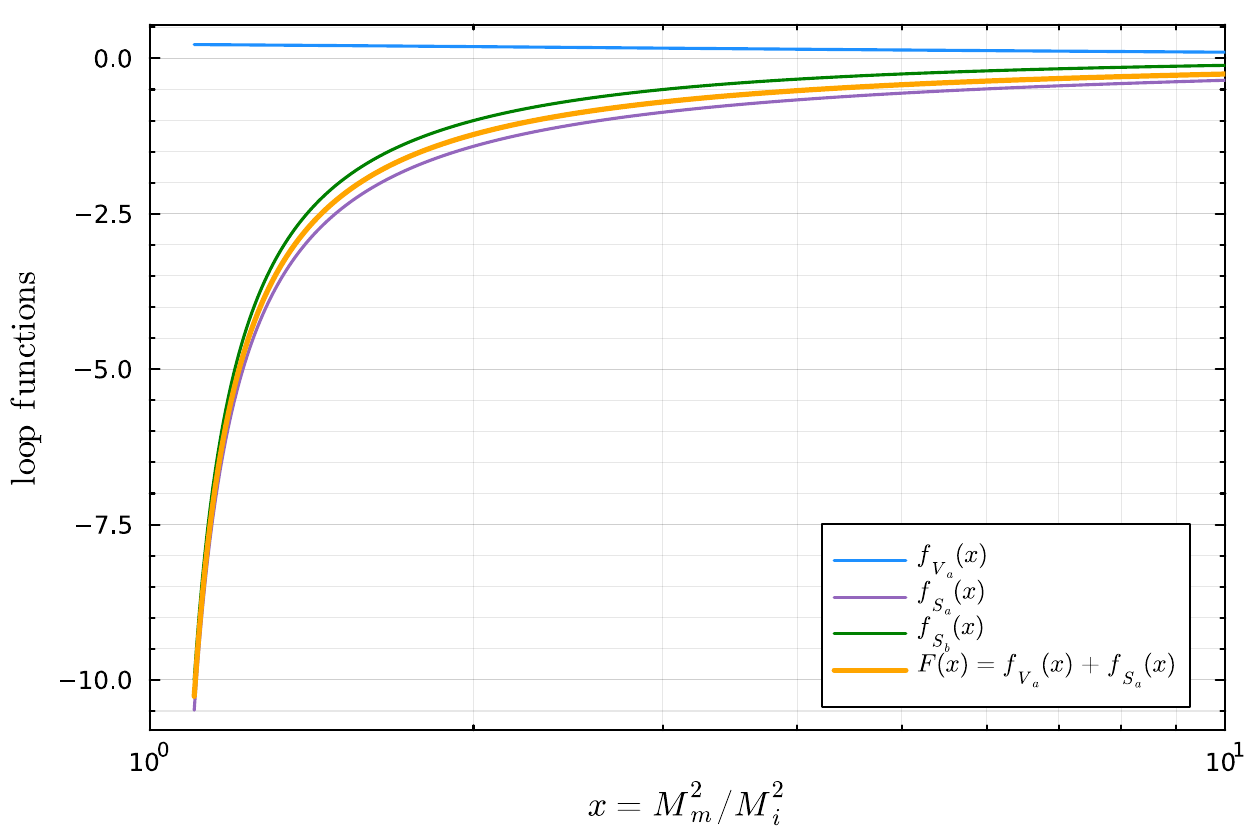}
\caption{Loop functions $f_{V_a}(x)$, $f_{S_a}(x)$, $f_{S_b}(x)$ and their sum $F(x)\coloneqq f_{V_a}(x)+f_{S_a}(x)$. In eq.~(\ref{eq:4.16}) the vertex and self-energy (a) enter as $(\mu^{\dagger}\mu)_{im}[f_{V_a}(x)+f_{S_a}(x)]$, while self-energy (b) enters as $(\mu^{\dagger}\mu)_{mi}f_{S_b}(x)$. After summing over lepton flavours, the $f_{S_b}(x)$ term vanishes (see eq.~(\ref{eq:4.17})); hence the effective loop factor is $F(x)=f_{V_a}(x)+f_{S_a}(x)$ for the flavour-summed asymmetry $\varepsilon_i=\sum_{\alpha}\varepsilon_{\alpha i}$. This form is not adopted in this work.}
\label{fig:6}
\end{figure}

Finally, we assemble the CP asymmetries. In our setup the dipole operator $\ONB$
dominates and it does not mix with other effective operators (see sec.~\ref{sec:3.2}). Consequently, the one-loop CP asymmetries inherit the same Lorentz structure as the tree-level widths; and the ratio of the $Z$-to-photon-channel asymmetries equals the tree-level phase-space/mixing-angle factor in eq.~(\ref{eq:4.5}):
\beq
\label{eq:4.15}
\dfrac{\varepsilon_{\alpha i}^Z}{\varepsilon_{\alpha i}^{\gamma}}
=(1-r_Z)^2(1+r_Z)\tan^2\theta_{\rm W}.
\eeq
Hence, the total CP asymmetries generated in the two-body decays of $N_i$ read
\beq
\label{eq:4.16}
\begin{split}
\varepsilon_{\alpha i}^{\gamma}+\varepsilon_{\alpha i}^Z
&=-\dfrac{M_i^2\cos^2\theta_{\rm W}}{2\pi\sum_{\beta}\abs{\mu_{\beta i}}^2}
\bqty*{1+(1-r_Z)^2(1+r_Z)\tan^2\theta_{\rm W}}\\[0.25em]
&\qquad\times\sum_{m\neq i}
\Im\bqty*{\mu_{\alpha i}^{\ast}\mu_{\alpha m}\Bqty*{(\mu^{\dagger}\mu)_{im}
[f_{V_a}(x)+f_{S_a}(x)]+(\mu^{\dagger}\mu)_{mi}f_{S_b}(x)}},
\end{split}
\eeq
where we used $\abs{\mu_{\beta i}^{\gamma}}^2+\abs{\mu_{\beta i}^Z}^2
=\abs{\mu_{\beta i}}^2(1+\tan^2\theta_{\rm W})
=\abs{\mu_{\beta i}}^2/\cos^2\theta_{\rm W}$ with $\mu_{\beta i}^{\gamma}\coloneqq\mu_{\beta i}$. The loop functions are those in eq.~(\ref{eq:4.14}) with $x=M_m^2/M_i^2$, and $r_Z=m_Z^2/M_i^2$.

Upon summing over lepton flavours one finds
\beq
\label{eq:4.17}
\sum_{\alpha}\Im\bqty*{(\mu^{\dagger}\mu)_{im}(\mu^{\dagger}\mu)_{mi}f_{S_b}(x)}
=\Im\bqty*{\abs{(\mu^{\dagger}\mu)_{im}}^2f_{S_b}(x)}=0,
\eeq
because $(\mu^{\dagger}\mu)_{im}(\mu^{\dagger}\mu)_{mi}=\abs{(\mu^{\dagger}\mu)_{im}}^2$ is real. Hence, $f_{S_b}(x)$ does not contribute to the flavour-summed asymmetry $\varepsilon_i=\sum_{\alpha}\varepsilon_{\alpha i}$. The effective loop factor is therefore
\beq
\label{eq:4.18}
F(x)\coloneqq f_{V_a}(x)+f_{S_a}(x),
\eeq
after flavour summation. In this work, we adopt the flavour-resolved asymmetries $\varepsilon_{\alpha i}$ ($\alpha=e,\mu,\tau$) and therefore $f_{S_b}(x)$ does not vanish.

\section{Kinetic framework: flavour and sphaleron dynamics}
\label{sec:5}

In what follows we work in the $N_1$-dominated approximation ($i=1$). In the $N_1$-dominated approximation, the heavier states $N_2,N_3,\ldots$\footnote{The antisymmetric structure in eq.~(\ref{eq:2.10}) requires at least two generations.} are integrated out and their CP phases enter $\varepsilon_{\alpha 1}$ only through $\sum_{m\neq 1}$. In this section we specify the kinetic setup used to evolve the Boltzmann system with four dynamical variables $\{Y_{N_1},Y_{\Delta e},Y_{\Delta\mu},Y_{\Delta\tau}\}$ ($Y_{\Delta\alpha}$ are defined in eq.~(\ref{eq:5.2})). We work in the fully-flavoured regime at electroweak temperatures, write down the flavoured Boltzmann equations (BE), summarise spectator effects through a flavour-coupling matrix, and state how the baryon asymmetry is obtained from leptonic charges via sphaleron conversion.

\subsection{Flavour regime at electroweak temperatures}
\label{sec:5.1}

At temperatures relevant to low-scale electromagnetic leptogenesis, $T=\mathcal{O}(100)~{\rm GeV}$, the charged-lepton Yukawa interactions are in equilibrium ($\varGamma_{\ell_{\alpha}}\gg H$). Consequently, flavour coherence is efficiently damped and the kinetic evolution is fully flavoured. We therefore track the three CP asymmetries $\{\varepsilon_{e1},\varepsilon_{\mu 1},\varepsilon_{\tau 1}\}$ separately and solve classical Boltzmann equations; a density-matrix description is not required because flavour decoherence is enforced by $\varGamma_{\ell_{\alpha}}\gg H$. This treatment is consistent with our $N_1$-dominated approximation: the heavier states $N_{m>1}$ are integrated out, and their CP-violating phases enter only through the loop-induced flavoured asymmetries $\varepsilon_{\alpha 1}$ which act as the sources in the Boltzmann system. Unless stated otherwise, we assume a thermal initial abundance for $N_1$ and zero initial asymmetries.

\subsection{Flavoured Boltzmann equations}
\label{sec:5.2}

We define $z\coloneqq M_1/T$ and work with the yield $Y_X\coloneqq n_X/s$, where $s$ denotes the entropy density. The kinetic equations read
\beq
\label{eq:5.1}
\begin{split}
\dv{Y_{N_1}}{z}
&=-D(z)(Y_{N_1}-Y_{N_1}^{\rm eq}),\\[0.25em]
\dv{Y_{\Delta\alpha}}{z}
&=\varepsilon_{\alpha 1}D(z)(Y_{N_1}-Y_{N_1}^{\rm eq})
-\sum_{\beta}W_{\alpha\beta}(z)Y_{\Delta\beta},
\end{split}
\eeq
where $Y_{\Delta\alpha}$ is the charge combination conserved by sphalerons,
\beq
\label{eq:5.2}
Y_{\Delta\alpha}\coloneqq\dfrac{1}{3}Y_B-Y_{L_{\alpha}},\qquad
\sum_{\alpha}Y_{\Delta\alpha}=Y_{B-L}.
\eeq
The decay parameter and the Hubble rate are defined as 
\beq
\label{eq:5.3}
K_1\coloneqq\dfrac{\varGamma_1}{H}\bigg|_{T=M_1},\qquad
H(z)=\dfrac{H_1}{z^2},\qquad
H_1\coloneqq H(T=M_1)=1.66\,\sqrt{\gstar}\,\dfrac{M_1^2}{M_{\rm Pl}},
\eeq
where $\varGamma_1$ is the two-body total width in eq.~(\ref{eq:4.8}) and $\gstar$ the effective number of relativistic degrees of freedom. In what follows we use $\Besselk_{\nu}(z)$ for the modified Bessel functions of the second kind; the symbol $K_1$ always denotes the decay parameter in eq.~(\ref{eq:5.3}). With these conventions, the decay factor $D(z)$ and the inverse-decay washout $W_{\rm ID}(z)$ are
\beq
\label{eq:5.4}
D(z)=K_1z\dfrac{\Besselk_1(z)}{\Besselk_2(z)},\qquad
W_{\rm ID}(z)=\dfrac{1}{4}K_1z^3\Besselk_1(z).
\eeq
To leading order we take
\beq
\label{eq:5.5}
W_{\alpha\beta}(z)\simeq P_{\alpha\beta}W_{\rm ID}(z),\qquad
P_{\alpha\beta}\simeq\diag(B_e,B_{\mu},B_{\tau}),\qquad
\sum_{\alpha}B_{\alpha}=1,
\eeq
where $B_{\alpha}$ are the flavour branching fractions of $N_1$ fixed by $\abs{C_{\alpha 1}}^2$ through $\varGamma_1$. Off-diagonal entries $P_{\alpha\neq\beta}$ from spectator effects can be included perturbatively; at the electroweak scale they are subleading numerically, and we set them to zero in our baseline prediction.

Using the widths in eqs.~(\ref{eq:4.6})--(\ref{eq:4.7}), the total two-body decay rate of $N_1$ can be written as
\beq
\label{eq:5.6}
\varGamma_1
=\sum_{\alpha}\varGamma_{\gamma,\alpha 1}^{\rm tree}
+\sum_{\alpha}\varGamma_{Z,\alpha 1}^{\rm tree}
=\dfrac{M_1^3}{2\pi}
\biggl(\sum_{\alpha}\abs{\mu_{\alpha 1}}^2\biggr)
R_Z(M_1),
\eeq
with
\beq
\label{eq:5.7}
R_Z(M_1)\coloneqq 1+(1-r_Z)^2(1+r_Z)\tan^2\theta_{\rm W},\qquad
r_Z=\dfrac{m_Z^2}{M_1^2}.
\eeq
We define the effective electromagnetic neutrino mass $\tilde{m}_1^{\rm EM}$ and the corresponding reference value $m_{\ast}^{\rm EM}$ as
\beq
\label{eq:5.8}
\tilde{m}_1^{\rm EM}\coloneqq v^2M_1\sum_{\alpha}\abs{\mu_{\alpha 1}}^2,
\qquad
m_{\ast}^{\rm EM}(M_1)\coloneqq
\dfrac{2\pi\cdot 1.66\,\sqrt{\gstar}\,v^2}{R_Z(M_1)\,M_{\rm Pl}},
\eeq
so that the decay parameter becomes\footnote{This coincides with the usual definition up to the $R_Z$ factor.}
\beq
\label{eq:5.9}
K_1\coloneqq\dfrac{\varGamma_1}{H}\bigg|_{T=M_1}
=\dfrac{\tilde{m}_1^{\rm EM}}{m_{\ast}^{\rm EM}(M_1)}
\simeq\dfrac{R_Z(M_1)}{5.35\times 10^{-4}~{\rm eV}}\,
\tilde{m}_1^{\rm EM}.
\eeq
In our benchmark (sec.~\ref{sec:6.1}), $K_1=1$ is realised for $\tilde{m}_1^{\rm EM}=m_{\ast}^{\rm EM}(M_1)\simeq 4.15\times 10^{-4}~{\rm eV}$.


\subsection{Spectator effects and the flavour-coupling matrix}
\label{sec:5.3}

Fast SM interactions (such as QCD and electroweak sphalerons and Yukawa scatterings) redistribute chemical potentials among particle species, thereby correlating the lepton-doublet asymmetries $Y_{\Delta\ell_{\alpha}}\coloneqq (n_{\ell_{\alpha}}-n_{\bar{\ell}_{\alpha}})/s$. In the fully-flavoured regime the redistribution among particle species is encoded by a flavour-coupling matrix $C^{\ell(3)}$ as~\cite{Nardi_2006, Abada_2006}
\beq
\label{eq:5.10}
Y_{\Delta\ell_{\alpha}}=\sum_{\beta}C_{\alpha\beta}^{\ell(3)}Y_{\Delta\beta},
\eeq
where $Y_{\Delta\ell_{\alpha}}$ should not be confused with $Y_{\Delta\alpha}$ defined in eq.~(\ref{eq:5.2}); the flavour-coupling matrix $C^{\ell(3)}$ is given by~\cite{ANTUSCH2012180}
\beq
\label{eq:5.11}
C^{\ell(3)}=\begin{pmatrix}
151/179&-20/179&-20/179\\-25/358&344/537&-14/537\\
-25/358&-14/537&344/537
\end{pmatrix}.
\eeq
This matrix is close to diagonal; including the off-diagonal entries only marginally changes the final asymmetry in our scans, and does not affect the qualitative conclusions. For clarity, we present baseline results in the diagonal approximation.

\subsection{Baryon conversion from sphalerons}
\label{sec:5.4}

In the broken phase relevant to our setup the final baryon asymmetry is obtained from the frozen $B-L$ charge as~\cite{PhysRevD.42.3344}
\beq
\label{eq:5.12}
Y_B=c_{\rm sph}\,Y_{B-L},\qquad
c_{\rm sph}=\dfrac{12}{37},
\eeq
whereas in the symmetric phase one would have $c_{\rm sph}=28/79$. Sphaleron processes switch off slightly below the electroweak symmetry breaking. In our numerics we assume that the baryon asymmetry is generated before this switch-off, and we evaluate $\YB$ at a fixed temperature $T\simeq T_{\rm sph}$ (small variations have negligible impact at our level of accuracy). The dependent dynamical variables are $\{Y_{N_1},Y_{\Delta e},Y_{\Delta\mu},Y_{\Delta\tau}\}$, consistent with the $N_1$-dominated approximation.

We model the electroweak window by the temperatures~\cite{PhysRevLett.113.141602, PhysRevD.93.025003}
\beq
\label{eq:5.13}
\begin{split}
T_{\rm EW}&\simeq 160~{\rm GeV}\quad
(\text{onset of the EWSB}),\\
T_{\rm sph}&\simeq 130~{\rm GeV}\quad
(\text{sphaleron freeze-out}),
\end{split}
\eeq
where charged-lepton Yukawa interactions are fully in equilibrium, so the three asymmetries $Y_{\Delta\alpha}$ decohere and evolve independently.

In our parameter range the washout is very weak, therefore the flavour asymmetries $Y_{\Delta\alpha}$ keep increasing slowly over the electroweak window and do not develop a clear late-time plateau (see fig.~\ref{fig:7}). By contrast, baryon-number conversion ceases abruptly once electroweak sphalerons switch off at $T\simeq T_{\rm sph}$. For this reason we define, in this work, the freeze-out baryon asymmetry at sphaleron decoupling,
\beq
\label{eq:5.14}
Y_B^{\rm FO}=c_{\rm sph}\,Y_{B-L}(T_{\rm sph}),\qquad
c_{\rm sph}=\dfrac{12}{37},
\eeq
i.e. we read out the asymmetries at sphaleron decoupling. This prescription avoids an arbitrary late-time choice and reflects the fact that only at temperatures above $T_{\rm sph}$ can lepton charges be converted into baryon number; the $N_1$ decays and inverse decays may still be active below $T_{\rm sph}$.

\section{Results}
\label{sec:6}

In this work, we neglect finite-temperature corrections to the dipole-induced decay widths and CP asymmetries; a systematic thermal-EFT treatment is left for future work. Concerning the washout, we account only for inverse decays, omitting $\Delta L=1$ and $\Delta L=2$ scatterings. These approximations are sufficient for the present analysis.

\subsection{Numerical setup}
\label{sec:6.1}

For the numerics, we assume two right-handed neutrinos $N_1$, $N_2$ and adopt the benchmark masses $(M_1,M_2,M_S,M_{\Psi})=(0.5,1.5,8,10)~{\rm TeV}$. We employ the $N_1$-dominated approximation. Moreover, we set $\mu_{\rm ref}=150~{\rm GeV}$ and take $\gstar=106.75$, $M_{\rm Pl}=1.22\times 10^{19}~{\rm GeV}$. Initial conditions are $Y_{\Delta\alpha}=0$ and a thermal abundance for $N_1$. Spectator effects are implemented through the $3\times 3$ flavour-coupling matrix $C^{\ell(3)}$ in eq.~(\ref{eq:5.11}), and branching fractions are collected in $P_{\alpha\beta}\simeq\diag(B_e,B_{\mu},B_{\tau})$ with $\sum_{\alpha}B_{\alpha}=1$. Baryon asymmetries are quoted as $Y_B=c_{\rm sph}\,Y_{B-L}$ with $c_{\rm sph}=12/37$ (sec.~\ref{sec:5.4}). 

Low-scale electromagnetic leptogenesis proceeds across the electroweak window from $T_{\rm EW}\simeq 160~{\rm GeV}$ down to sphaleron freeze-out at $T_{\rm sph}\simeq 130~{\rm GeV}$. All inputs are specified at $\mu_{\rm ref}=150~{\rm GeV}$, with gauge couplings and the top Yukawa evolved down to this scale as in sec.~\ref{sec:3}. As explained in sec.~\ref{sec:5.4}, we define the freeze-out of the asymmetries to occur at sphaleron decoupling $\YB^{\rm FO}=c_{\rm sph}\,Y_{B-L}(T_{\rm sph})$ in this work, because the $Y_{\Delta\alpha}$ curves do not form a clear plateau (cf. fig.~\ref{fig:7}).

\subsection{Solutions of Boltzmann equations}
\label{sec:6.2}

\begin{figure}[t]
\centering
\includegraphics[width=\linewidth]{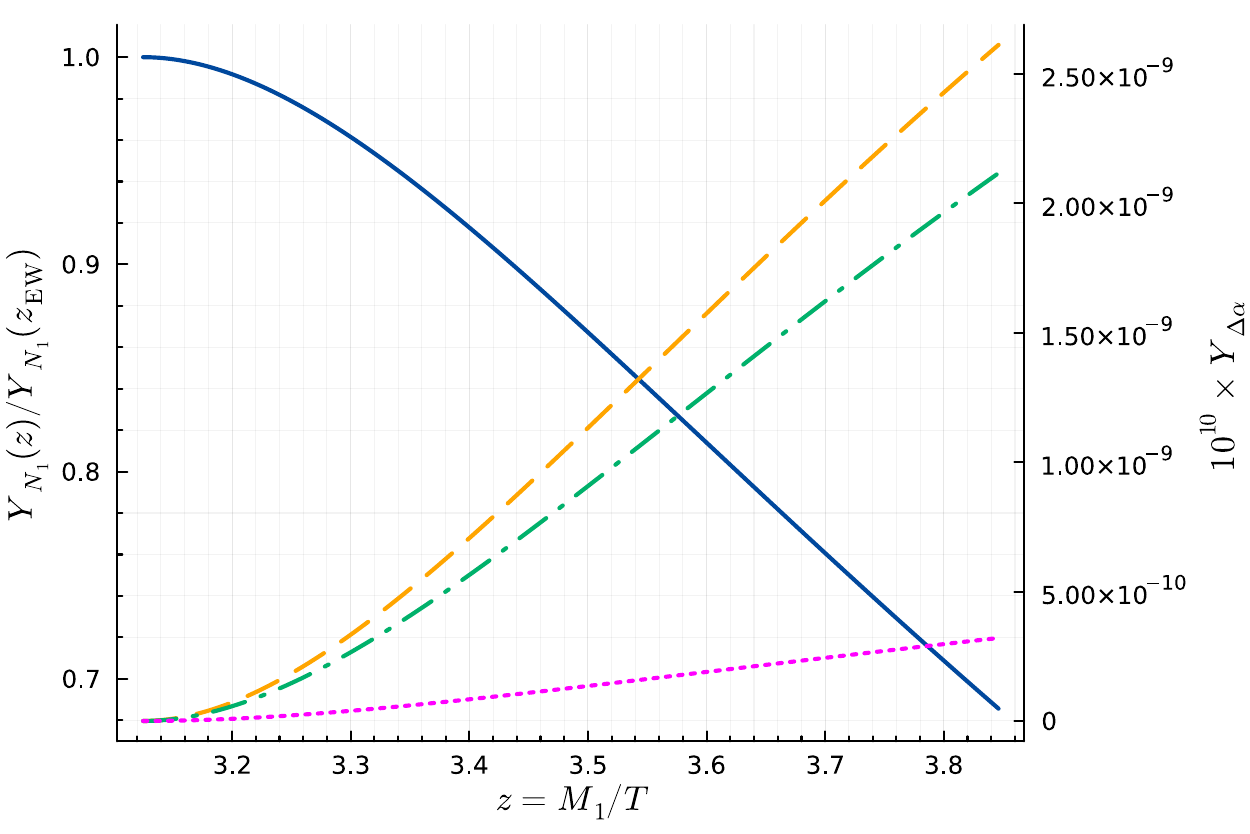}
\caption{Time evolution of the $N_1$ abundance and of the flavoured lepton asymmetries in the fully-flavoured regime, shown here for $\tilde{m}_1^{\rm EM}=10^{-3}~{\rm eV}$. The time window spans from $T_{\rm EW}\simeq 160~{\rm GeV}$ (left end) to $T_{\rm sph}\simeq 130~{\rm GeV}$ (right end). Plotted as functions of $z=M_1/T$: left axis---the normalised $N_1$ abundance $Y_{N_1}(z)/Y_{N_1}(z_{\rm EW})$ (blue), with $z_{\rm EW}=M_1/T_{\rm EW}$; right axis---the flavour charges $Y_{\Delta e}$ (orange, dashed), $Y_{\Delta\mu}$ (green, dash-dotted) and $Y_{\Delta\tau}$ (magenta, dotted), rescaled as $10^{10}\,Y_{\Delta\alpha}$. The asymmetries are sourced when $N_1$ departs from equilibrium and freeze out once electroweak sphalerons decouple at $T\simeq T_{\rm sph}$. All inputs are as specified in sec.~\ref{sec:6.1}. We extract $\YB^{\rm FO}$ from the flavour-resolved solutions at $T\simeq T_{\rm sph}$.}
\label{fig:7}
\end{figure}


In all plots and numbers quoted in this section we solve the three-flavour Boltzmann system in eq.~(\ref{eq:5.1})
and read out the freeze-out baryon asymmetry at sphaleron decoupling (cf. eq.~(\ref{eq:5.14})),
\beq
\label{eq:6.1}
\YB^{\rm FO}=c_{\rm sph}\sum_{\alpha}Y_{\Delta\alpha}(T_{\rm sph}),\qquad
c_{\rm sph}=\dfrac{12}{37}.
\eeq
This removes any ambiguity related to flavour averaging and keeps track of spectator-induced flavour correlations directly.

Fig.~\ref{fig:7} shows a representative evolution as a function of $z=M_1/T$. In the fully flavoured Boltzmann system, the source term is proportional to $\varepsilon_{\alpha 1}D(z)(Y_{N_1}-Y_{N_1}^{\rm eq})$, whereas the washout term is governed by $\sum_{\beta}W_{\alpha\beta}(z)Y_{\Delta\beta}$ (see sec.~\ref{sec:5.2}). After a short transient the charged-lepton Yukawas fully decohere the flavours, and the three asymmetries $\{Y_{\Delta e},Y_{\Delta\mu},Y_{\Delta\tau}\}$ evolve separately. The asymmetries grow as $N_1$ departs from equilibrium, remain tiny, and are finally read out at sphaleron freeze-out ($T\simeq T_{\rm sph}$), when electroweak sphaleron transitions cease to convert $B-L$ into $B$.

Within the broken-phase window $z\in[z_{\rm EW},z_{\rm sph}]$, the washout kernel $W_{\alpha\beta}(z)$ remains subleading while the weight function $w(z)\coloneqq D(z)[Y_{N_1}(z)-Y_{N_1}^{\rm eq}(z)]$ monotonically grows as $N_1$ departs from equilibrium. The integral $Y_{\Delta\alpha}(z)\sim\int^{z}\dd z'\;\varepsilon_{\alpha 1}w(z')$ is therefore dominated towards the end of the window, explaining the steady rise of all three flavour asymmetries until sphaleron decoupling at $T\simeq T_{\rm sph}$.

\subsection{Abundance of the final baryon asymmetry}
\label{sec:6.3}

\begin{figure}[t]
\centering
\includegraphics[width=\linewidth]{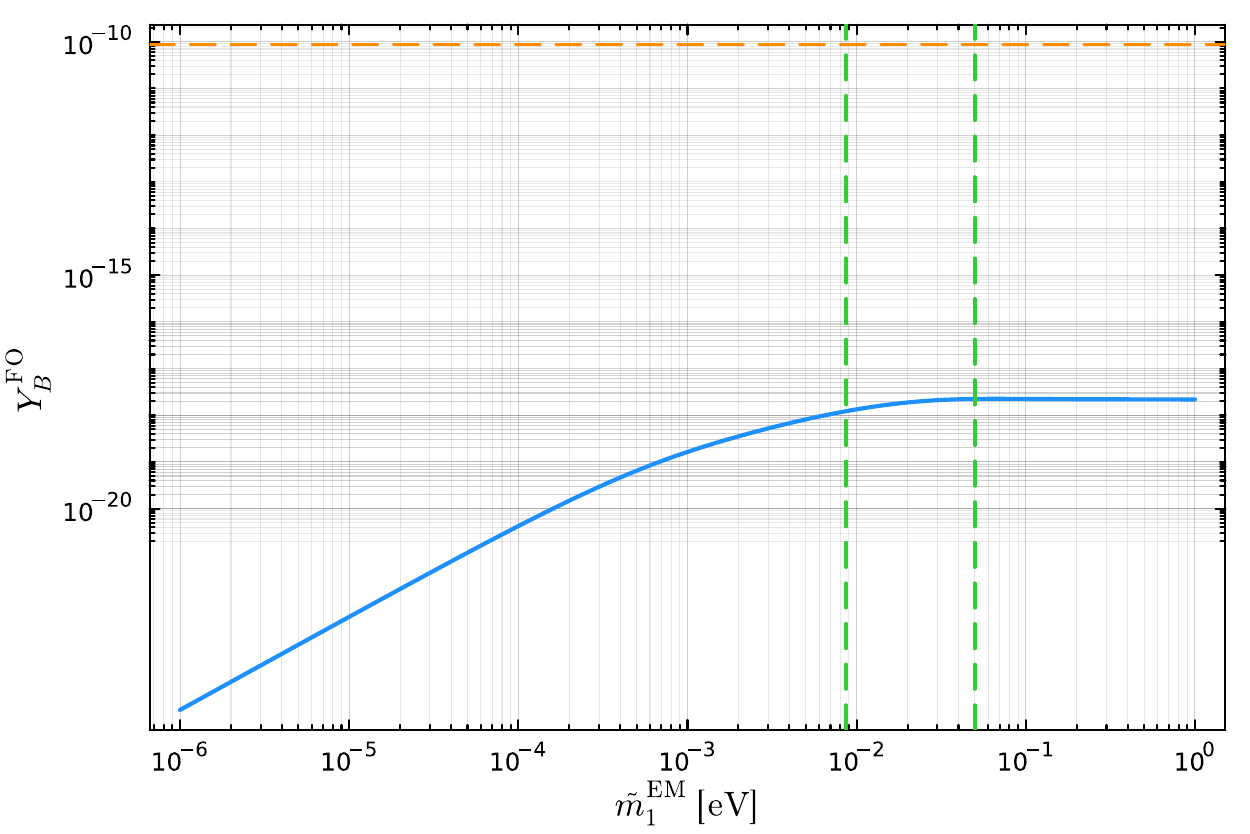}
\caption{Freeze-out baryon asymmetry $Y_B^{\rm FO}$ as a function of
$\tilde{m}_1^{\rm EM}$ (blue).
The dashed orange line indicates the observed value $\YB^{\rm obs}\simeq 8.7\times 10^{-11}$~\cite{2020A&A...641A...6P}. The dashed green band indicates the light-neutrino mass scale suggested by oscillation data~\cite{Esteban:2024eli}. The read-out is performed at $T_{\rm sph}\simeq 130~{\rm GeV}$ as in fig.~\ref{fig:7}. In the weak-washout regime ($\tilde{m}_1^{\rm EM}\lesssim 4.15\times 10^{-4}~{\rm eV}$, see eq.~(\ref{eq:5.9})), the monotonic rise of $\YB^{\rm FO}$ with $\tilde{m}_1^{\rm EM}$ simply tracks the increase of the source.
In the strong-washout regime, however, the simultaneous growth of the washout offsets the gain, preventing $\YB^{\rm FO}$ from approaching $\YB^{\rm obs}$; hence, the asymmetry forms a plateau of height $\cO(10^{-18}\text{--}10^{-17})$.}
\label{fig:8}
\end{figure}

Fig.~\ref{fig:8} displays $\YB^{\rm FO}$ as a function of $\tilde{m}_1^{\rm EM}$. From eq.~(\ref{eq:5.9}) it follows that
\beq
\label{eq:6.2}
\tilde{m}_1^{\rm EM}
=m_{\ast}^{\rm EM}(M_1)\,K_1
=\dfrac{5.35\times 10^{-4}~{\rm eV}}{R_Z(M_1)}\,K_1,
\eeq
where $R_Z(M_1)$ is defined in eq.~(\ref{eq:5.7}). The result increases monotonically with $\tilde{m}_1^{\rm EM}$ but remains at least seven to eight orders of magnitude below the observed value $\YB^{\rm obs}\simeq 8.7\times 10^{-11}$~\cite{2020A&A...641A...6P}.
This behaviour reflects a structural scaling. The CP sources $\varepsilon_{\alpha 1}$ in eq.~(\ref{eq:4.16}) are quadratic in the dipole couplings $\mu_{\alpha 1}$, while the washout strength is controlled by $K_1\propto\sum_{\alpha}\abs{\mu_{\alpha 1}}^2$ (eqs.~(\ref{eq:5.8})--(\ref{eq:5.9})). 
At fixed dimensionless heavy-mass ratios $r=M_S^2/M_{\Psi}^2$ and $\xi_j=M_{\Psi}/M_j$ in eq.~(\ref{eq:2.10}), the heavy-scale matching gives $\mu\propto v/M_{\Psi}^2$. One then finds the following parametric scaling:
\beq
\label{eq:6.3}
\varepsilon_{\alpha 1}\propto M_{\Psi}^{-4},\qquad
K_1\propto M_{\Psi}^{-4}.
\eeq
In either language, strengthening the dipole interaction increases production and washout in lockstep, leaving the final asymmetry strongly suppressed in the non-resonant regime.

The dashed green band in fig.~\ref{fig:8} indicates the characteristic light-neutrino mass scale suggested by oscillation data. Concretely, we adopt the  NuFIT 6.0 global fit in normal ordering, whose best-fit mass-squared splittings are~\cite{Esteban:2024eli}
\beq
\label{eq:6.4}
\Delta m_{21}^2=7.49\times 10^{-5}~{\rm eV}^2,\qquad
\Delta m_{31}^2=2.513\times 10^{-3}~{\rm eV}^2,
\eeq
so that in the limit $m_1\to 0$ one has $m_2=\sqrt{\Delta m_{21}^2}\simeq 8.65~{\rm meV}$, $m_3=\sqrt{\Delta m_{31}^2}\simeq 50.1~{\rm meV}$.
\section{Discussion and outlook}
\label{sec:7}

\subsection{Why the signal is suppressed}
\label{sec:7.1}

\enlargethispage{\baselineskip}

Our computation follows a single, consistent chain: UV theory $\;\longrightarrow\;$ one-loop matching $\;\longrightarrow\;$ one-loop running $\;\longrightarrow\;$ dipole couplings $\mu_{\alpha i}$ $\;\longrightarrow\;$ decay widths and CP asymmetries $\;\longrightarrow\;$ fully flavoured Boltzmann system. Numerically (see figs.~\ref{fig:7}--\ref{fig:8}), low-scale electromagnetic leptogenesis (EMLG) yields a freeze-out baryon asymmetry
\beq
\label{eq:7.1}
Y_B^{\rm FO}\lesssim 10^{-17},
\eeq
i.e. seven to eight orders of magnitude below the observed value $Y_B^{\rm obs}\simeq 8.7\times 10^{-11}$~\cite{2020A&A...641A...6P}.

The origin of this suppression is structural. Gauge invariance forces the electromagnetic dipole to arise from the dimension-six operator
\beq
\label{eq:7.2}
\ONB=(\bar{L}\sigma^{\mu\nu}N)\tilde{H}B_{\mu\nu},
\eeq
with a Higgs insertion. After electroweak symmetry breaking (EWSB) the gauge-invariant operator matches onto the broken-phase dipole coupling,
\beq
\label{eq:7.3}
\mu_{\alpha i}=\dfrac{v\cos\theta_{\rm W}}{\sqrt{2}\,M_{\Psi}^2}\,
C_{NB,\alpha i}(\mu_{\rm ref}),
\eeq
and the widths and CP asymmetries are as in sec.~\ref{sec:4}. We keep the CP asymmetries flavour-resolved, i.e. we work with $\varepsilon_{\alpha i}$ rather than with the flavour-summed combination.
Furthermore, we keep fixed the dimensionless heavy-mass ratios $r=M_S^2/M_{\Psi}^2$ and $\xi_j=M_{\Psi}/M_j$ that parametrise the one-loop matching. Their effect is absorbed into $\cO(1)$ loop functions and does not affect the power counting in $M_{\Psi}$. Since the dipole couplings scale as $\mu_{\alpha i}\propto v/M_{\Psi}^2$, both the total two-body widths and the flavour-resolved CP asymmetries inherit the common parametric suppression
\beq
\label{eq:7.4}
\varGamma_i,\;\varepsilon_{\alpha i}\propto
\dfrac{v^2}{M_{\Psi}^4},\quad
(\text{up to loop and RGE factors}).
\eeq
This common quadratic scaling in $\mu$ explains the persistent suppression in the non-resonant, hierarchical regime seen in figs.~\ref{fig:7}--\ref{fig:8}.

\subsection{Relation to recent work}
\label{sec:7.2}


Below we briefly comment on a variant of EMLG that employs a right-right dipole. In ref.~\cite{BORAH2025139557}, Borah--Dasgupta introduced a dimension-five dipole operator between a heavy $N_i$ and a light sterile $\nu_R$, $\frac{\lambda_{i\alpha}}{\Lambda}\bar{N}_i\sigma^{\mu\nu}\nu_{R\alpha}B_{\mu\nu}$. The heavy $N_i$ then undergoes two-body decays $N_i\to\nu_R+B_{\mu}$, and a CP asymmetry arises at one loop from the interference with graphs containing a second heavy state $N_j$. The asymmetry produced in the $\nu_R$ sector must subsequently be transferred to left-handed leptons via Yukawa interaction with a neutrinophilic Higgs so that electroweak sphalerons can convert it into a baryon asymmetry; the same couplings also radiatively induce a Majorana mass for $\nu_R$, further constraining the viable parameter space.

Furthermore, they have proposed a cogenesis scenario in which heavy vector-like fermions $N$ decay via a dimension-five dipole operator into the right-chiral component $\nu_R$ of a Dirac neutrino and a dark fermion $\chi$~\cite{Borah:2025dka}. CP asymmetries are generated in the $\nu_R$ and $\chi$ sectors such that $\varepsilon_{\nu_R}+\varepsilon_{\chi_L}+\varepsilon_{\chi_R}=0$. The $\nu_R$ asymmetry is transferred to the SM lepton doublets via Yukawa interactions with a neutrinophilic Higgs doublet before sphaleron freeze-out. This links the baryon asymmetry to an asymmetric DM component, implying $m_{\chi}\sim\cO(1)~{\rm GeV}$ and yielding distinctive signatures such as $\chi\to\gamma\gamma\nu_R$, as well as a contribution $\Delta N_{\rm eff}\simeq 0.14$ from thermalised $\nu_R$. A viable region in the $(\Lambda,M_1)$ plane emerges after imposing constraints from washout, lifetimes, and indirect detection.


By contrast, our EFT-based analysis of EMLG can be viewed as a concrete realisation of the weak-washout, decay-driven class of leptogenesis, with the same \enquote{mass-coupling cancellation} mechanism emphasised by Kanemura--Li~\cite{Kanemura:2025rsy}. In our subset, the decay coupling is an effective dipole fixed by one-loop matching rather than a free Yukawa; hence any uniform strengthening of the underlying couplings feeds into $\mu$ and enhances both the CP source and the washout in lockstep ($\varepsilon_{\alpha i},\;K_i\propto\mu^2$), leaving the final lepton asymmetry, and hence the baryon asymmetry, strongly suppressed.\footnote{Here \enquote{cancellation} refers to the $\mu^2$ co-scaling of production and washout (not a direct cancellation between a mass and a coupling).}

\subsection{Relation to direct searches}
\label{sec:7.3}

Collider and intensity-frontier experiments provide complementary probes of the dipole interactions relevant to the EMLG setup~\cite{Abdullahi_2023}. Effective dipole interactions of heavy neutral leptons (HNLs) have recently been investigated in dedicated search strategies. Barducci et al. derived LHC sensitivities to RHN transition-dipole operators in channels with $\tau$-leptons, and mapped the reach in the corresponding EFT parameter space~\cite{Barducci:2022gdv}. In a follow-up they proposed a mono-$\gamma$ $+$ missing-energy ($N\to\nu\gamma$) search at NA62 that directly targets dipole-induced HNL production and decay~\cite{Barducci:2024nvd}. Our EFT and low-scale setup are complementary: while we compute the leptogenesis dynamics and the RG effects that control $\CNB$, these studies quantify current and near-term experimental reach on the very same dipole structures.

\subsection{A plausible way forward: resonant EMLG}
\label{sec:7.4}

The analysis in secs.~\ref{sec:5}--\ref{sec:6} assumed a hierarchical spectrum, $x\gtrsim 10$.\footnote{In the benchmark of sec.~\ref{sec:6} we use $x=M_2^2/M_1^2=9$. As seen from fig.~\ref{fig:6}, the loop functions are already very close to their $\cO(1)$ asymptotics, so the \enquote{hierarchical} statements apply unchanged.} By contrast, in the quasi-degenerate limit $x\to 1$ the loop functions in fig.~\ref{fig:6} develop the familiar resonant behaviour.
For flavour-summed asymmetries, the $f_{S_b}(x)$ piece cancels after summing over $\alpha$ (eq.~(\ref{eq:4.17})). However, if one does not perform the flavour sum---i.e. one follows the three-flavour asymmetries $\{Y_{\Delta e},Y_{\Delta\mu},Y_{\Delta\tau}\}$ separately as in our kinetic setup---both resonant sources $f_{S_a}(x)$ and $f_{S_b}(x)$ can survive and contribute.

This observation suggests a realistic avenue: arrange a small mass splitting $\Delta M\coloneqq M_m-M_1\ll M_1$ so that the loop functions of self-energy contributions resonate while keeping the EFT under quantitative control. A dedicated analysis in the quasi-degenerate regime, with full flavour resolution in the Boltzmann (or density matrix) dynamics and a proper resummation of the self-energy denominators, is therefore a promising direction to reach $\YB^{\rm obs}$.

\section{Conclusions}
\label{sec:8}

In summary, our low-scale analysis leads to the following points.
\begin{itemize}
\item Structural suppression. Gauge invariance of the effective operator enforces a Higgs insertion, $\mu\propto v/M_{\Psi}^2$. As a result, both $\varGamma_i$ and $\varepsilon_{\alpha i}$ scale as $v^2/M_{\Psi}^4$ and carry an additional loop suppression from the one-loop matching (via $\CNB$) and the subsequent RGE.

\item Hierarchical regime. In the $N_1$-dominated, hierarchical case $x\gtrsim 10$ the flavour-resolved loop factors are $\cO(1)$, leaving $\YB^{\rm FO}\lesssim 10^{-17}$.


\item Resonant window. In the quasi-degenerate setup $x\to 1$ the self-energy enhancement re-emerges. With flavour separation the $f_{S_b}$ term does not cancel and can dominate, opening a credible path to $Y_B^{\rm obs}$.
\end{itemize}

These conclusions sharpen the targets for future work: (i) a controlled resonant analysis with flavour-resolved kinetics (Boltzmann or matrix of densities) with self-energy resummation; and (ii) UV constructions that mitigate the Higgs-insertion suppression while preserving gauge invariance.

\acknowledgments

The author is grateful to R.~Jinno for helpful comments on the manuscript. The author also thanks D.~Borah and A.~Dasgupta for informing her of recent related paper~\cite{Borah:2025dka}.

\appendix
\section{One-loop matching for $\ONB$ and $\CNB$}
\label{sec:A}

\subsection{Full-theory 1PI vertex $\langle L_{\alpha}B_{\mu}\tilde{H}N_i\rangle$ at one loop}
\label{sec:A.1}

We work in dimensional regularisation with $n=4-2\epsilon$ and 't Hooft--Feynman gauge ($\xi=1$). We denote the external $\mathrm{U}(1)_Y$ momentum by $q$, and expand the numerator to first order in $q$ in the regime $q^2\ll M_{\Psi}^2$.

Of the two vertex topologies, it suffices to evaluate the graph in which the $\mathrm{U}(1)_Y$ gauge boson is emitted from the fermion line (fig.~\ref{fig:2}); we then take the antisymmetric combination under $i\leftrightarrow j$. Introducing Feynman parameters, one obtains
\begin{align}
\mathrm{i}\mathcal{M}_{\rm full}
&=(\mathrm{i}g_1Y_{\Psi})(-\lambda_i\lambda'_j)(-\mathrm{i}y_{\alpha j})
\bar{u}_{L_{\alpha}}\dfrac{\mathrm{i}}{\slashed{p}-M_j}P_R\notag\\
&\quad\times
\int\!\!\dfrac{\dd^4k}{(2\pi)^4}\;
\dfrac{P_R(\slashed{k}+M_{\Psi})\gamma^{\mu}
(\slashed{k}+\slashed{q}+M_{\Psi})P_R}
{(k^2-M_{\Psi}^2)\bqty*{(k+q)^2-M_{\Psi}^2}\bqty*{(k+q)^2-M_S^2}}\,
u_{N_i}\varepsilon_{\mu}
\notag\\
&=\mathrm{i}g_1Y_{\Psi}\lambda_i\lambda'_j\dfrac{y_{\alpha j}}{M_j}
\int\!\!\dfrac{\dd^4k}{(2\pi)^4}\int_{0}^{1}\dd x\int_{0}^{1-x}\dd y\notag\\
&\quad\times
\dfrac{2\bar{u}_{L_{\alpha}}(\slashed{k}+M_{\Psi})\gamma^{\mu}
(\slashed{k}+\slashed{q}+M_{\Psi})P_Ru_{N_i}\varepsilon_{\mu}}
{\{(k^2-M_{\Psi}^2)x+\bqty*{(k+q)^2-M_{\Psi}^2}y
+\bqty*{(k+q)^2-M_S^2}(1-x-y)\}^3}.
\label{eq:A.1}
\end{align}
We work in the kinematic regime $p^2\ll M_j^2$ and expand the numerator to first order in the external $\mathrm{U}(1)_Y$ gauge-boson momentum $q$:
\begin{align}
(\slashed{k}+M_{\Psi})\gamma^{\mu}(\slashed{k}+\slashed{q}+M_{\Psi})
=(\slashed{k}+M_{\Psi})\gamma^{\mu}(\slashed{k}+M_{\Psi})
+(\slashed{k}+M_{\Psi})\gamma^{\mu}\slashed{q}.
\label{eq:A.2}
\end{align}
The first term does not contribute to a magnetic form factor and cancels against the mirror graph by the Ward--Takahashi identity. The magnetic dipole piece comes from the second term. Substituting it into~(\ref{eq:A.1}) and taking the on-shell $\mathrm{U}(1)_Y$ gauge-boson limit $q^2\to 0$ at the end, the loop integration in dimensional regularisation yields
\begin{align}
\mathrm{i}\mathcal{M}_{\rm full}
&=\mathrm{i}g_1Y_{\Psi}\lambda_i\lambda'_j\dfrac{y_{\alpha j}}{M_j}
\int_{0}^{1}\dd x\int_{0}^{1-x}\dd y\;
\int\!\!\dfrac{\dd^nk}{(2\pi)^n}\notag\\
&\quad\times
\dfrac{2\bar{u}_{L_{\alpha}}(\slashed{k}+M_{\Psi})
\gamma^{\mu}\slashed{q}P_Ru_{N_i}\varepsilon_{\mu}}
{[M_{\Psi}^2(x+y)+M_S^2(1-x-y)-q^2x(1-x)+2k\cdot q(x-1)-k^2]^3}\notag\\
&=\mathrm{i}g_1Y_{\Psi}\lambda_i\lambda'_j\dfrac{y_{\alpha j}}{M_j}
\int_{0}^{1}\dd x\int_{0}^{1-x}\dd y\notag\\
&\quad\times
\dfrac{\mathrm{i}\varGamma\pqty*{3-\frac{n}{2}}}
{(4\pi)^{\frac{n}{2}}\varGamma(3)}
\dfrac{2\bar{u}_{L_{\alpha}}[\slashed{q}(x-1)+M_{\Psi}]
\gamma^{\mu}\slashed{q}P_Ru_{N_i}\varepsilon_{\mu}}
{[M_{\Psi}^2(x+y)+M_S^2(1-x-y)-q^2x(1-x)]^{3-\frac{n}{2}}}\notag\\
&=-g_1Y_{\Psi}\lambda_i\lambda'_j\dfrac{y_{\alpha j}}{M_j}
\int_{0}^{1}\dd x\int_{0}^{1-x}\dd y\;
\dfrac{1}{16\pi^2}
\dfrac{\bar{u}_{L_{\alpha}}
(M_{\Psi}\gamma^{\mu}\slashed{q})P_Ru_{N_i}}
{M_{\Psi}^2(x+y)+M_S^2(1-x-y)}.
\label{eq:A.3}
\end{align}
To isolate the magnetic structure we use the Clifford-algebra identity
\beq
\label{eq:A.4}
\gamma^{\mu}\slashed{q}=q^{\mu}-\mathrm{i}\sigma^{\mu\nu}q_{\nu},\qquad
\sigma^{\mu\nu}\coloneqq\dfrac{\mathrm{i}}{2}[\gamma^{\mu},\gamma^{\nu}],
\eeq
which, in the presence of the loop numerator $M_{\Psi}\gamma^{\mu}\slashed{q}$, implies 
\beq
\label{eq:A.5}
M_{\Psi}\gamma^{\mu}\slashed{q}\;\longrightarrow\;
M_{\Psi}(q^{\mu}-\mathrm{i}\sigma^{\mu\nu}q_{\nu}).
\eeq
The electric piece proportional to $q^{\mu}$ vanishes in the sum of the two vertex graphs by the Ward--Takahashi identity; hence only the magnetic structure $-\mathrm{i}\sigma^{\mu\nu}q_{\nu}$ can survive. 

\enlargethispage{2\baselineskip}

It is convenient to define $r\coloneqq M_S^2/M_{\Psi}^2$. The remaining Feynman-parameter integral reduces to
\begin{align}
\mathrm{i}\mathcal{M}_{\rm full}
&=\mathrm{i}g_1Y_{\Psi}\lambda_i\lambda'_j\dfrac{y_{\alpha j}}{M_j}
\int_{0}^{1}\dd x\int_{0}^{1-x}\dd y\;
\dfrac{1}{16\pi^2}
\dfrac{\bar{u}_{L_{\alpha}}
(M_{\Psi}\sigma^{\mu\nu}q_{\nu})P_Ru_{N_i}\varepsilon_{\mu}}
{M_{\Psi}^2(x+y)+M_S^2(1-x-y)}\notag\\
&=\dfrac{\mathrm{i}g_1Y_{\Psi}\lambda_i\lambda'_j}{16\pi^2M_{\Psi}}\cdot
\dfrac{y_{\alpha j}}{M_j}
\int_{0}^{1}\dd x\int_{0}^{1-x}\dd y\;
\dfrac{\bar{u}_{L_{\alpha}}\sigma^{\mu\nu}q_{\nu}P_Ru_{N_i}\varepsilon_{\mu}}
{(1-r)(x+y)+r}.
\label{eq:A.6}
\end{align}
Performing the parameter integrals over the unit triangle gives the standard loop function\footnote{For the loop function $g(r)=(1-r+r\ln r)/(1-r)^2$, the limiting behaviours are: $g(r\to 1)\to 1/2$, $g(r\ll 1)\simeq 1+\ln r$, and $g(r\gg 1)\simeq \ln r/r$. Hence, one finds that no singularity arises from the $(1-r)^2$ in the denominator; matching the smooth behaviour of $g(r)$ used in eq.~(\ref{eq:2.10}).}
\beq
\label{eq:A.7}
g(r)\coloneqq\int_{0}^{1}\dfrac{z}{(1-r)z+r}\;\dd z
=\dfrac{1}{1-r}+\dfrac{r\ln r}{(1-r)^2},\qquad
r=\dfrac{M_S^2}{M_{\Psi}^2}.
\eeq
Adding the two vertex graphs, one obtains the full-theory 1PI vertex
\beq
\label{eq:A.8}
\mathcal{M}_{\rm full}
=\sum_{j}\dfrac{g_1Y_{\Psi}}{16\pi^2M_{\Psi}}
(\lambda_i\lambda'_j-\lambda_j\lambda'_i)
\dfrac{y_{\alpha j}}{M_j}\cdot
\dfrac{1-r+r\ln r}{(1-r)^2}\cdot
\bar{u}_{L_{\alpha}}\sigma^{\mu\nu}q_{\nu}P_Ru_{N_i}\varepsilon_{\mu},
\eeq
which is antisymmetric in $i\leftrightarrow j$ as required by the Majorana nature of $N_i$. The Lorentz structure is thus $\bar{L}_{\alpha}\sigma^{\mu\nu}P_RN_iq_{\nu}\varepsilon_{\mu}$. In the symmetric phase this Lorentz structure is reproduced by the gauge-invariant operator\footnote{The identity~(\ref{eq:A.4}) is used solely to isolate the magnetic piece. The \enquote{electric} contribution proportional to $q^{\mu}$ cancels between the two vertex graphs by current conservation (Ward--Takahashi identity), which is why only the $\sigma^{\mu\nu}q_{\nu}$ structure survives in~(\ref{eq:A.9}).}
\beq
\label{eq:A.9}
\cO_{NB,\alpha i}=(\bar{L}_{\alpha}\sigma^{\mu\nu}P_RN_i)\tilde{H}B_{\mu\nu},
\eeq
where the insertion of $\tilde{H}$ is mandated by gauge invariance, see also sec.~\ref{sec:2}.

\subsection{Matching to the EFT coefficient at $\mu=M_{\Psi}$}
\label{sec:A.2}

We use the dimensionless normalisation of $\CNB$ adopted in the main text; see eq.~(\ref{eq:2.8}). In the EFT we introduce
\beq
\label{eq:A.10}
-\mathcal{L}_{\rm EFT}\supset\dfrac{C_{NB,\alpha i}^{\rm EFT}(\mu)}{\Lambda^2}\,\cO_{NB,\alpha i}+\text{h.c.},
\eeq
and evaluate the tree-level 1PI vertex for the same external legs $\{L_{\alpha},B_{\mu},\tilde{H},N_i\}$,\footnote{We match amputated 1PI Green functions with fixed external quantum numbers at momenta $\ll M_{\Psi}$. Setting $q_H=0$ projects onto the dimension-six dipole without higher-derivative operators. No on-shell process is implied; by crossing symmetry the result is channel-independent.}
\beq
\label{eq:A.11}
\mathcal{M}_{\rm EFT}
=\dfrac{C_{NB,\alpha i}^{\rm EFT}(\mu)}{\Lambda^2}\,
\bar{u}_{L_{\alpha}}(k_L)\sigma^{\mu\nu}q_{\nu}P_Ru_{N_i}(k)\varepsilon_{\mu}\tilde{H}(q_H),\qquad
q_H=0.
\eeq
Equating the amputated tree-level EFT and one-loop full-theory 1PI vertices at the heavy scale $\mu=\Lambda=M_{\Psi}$, 
\beq
\label{eq:A.12}
\mathcal{M}_{\rm EFT}^{\rm tree}(k'\ll M_{\Psi})
=\mathcal{M}_{\rm full}^{\text{\scriptsize 1-loop}}(k'\ll M_{\Psi}),
\eeq
gives
\beq
\label{eq:A.13}
C_{NB,\alpha i}^{\rm EFT}(M_{\Psi})=C_{NB,\alpha i}^{\rm full}(M_{\Psi}),
\eeq
with
\beq
\label{eq:A.14}
C_{NB,\alpha i}^{\rm full}(M_{\Psi})
=\sum_{j}\dfrac{g_1Y_{\Psi}}{16\pi^2}
(\lambda_i\lambda'_j-\lambda_j\lambda'_i)y_{\alpha j}
\dfrac{M_{\Psi}}{M_j}\cdot
\dfrac{1-r+r\ln r}{(1-r)^2},\qquad
r=\dfrac{M_S^2}{M_{\Psi}^2}.
\eeq
In the EFT the heavy fields $\Psi$ and $S$ are integrated out, while the right-handed neutrinos $N_{m>1}$ remain dynamical prior to imposing the $N_1$-dominated approximation used in the phenomenological sections. 

\section{One-loop renormalisation of $\ONB$ and $\CNB$}
\label{sec:B}

Throughout this appendix we work in the symmetric phase and set $m_{\ell}=0$, $m_H=0$. To avoid spurious infrared sensitivity we keep the external momentum off shell, $q^2\neq 0$; the $\overline{\rm MS}$ UV poles---and hence the anomalous dimensions---are independent of this choice. We adopt dimensional regularisation in $n=4-2\epsilon$ with the $\overline{\rm MS}$ scheme, and we fix the gauge to the 't Hooft--Feynman choice $\xi=1$.
We extract UV poles as $\bar{\epsilon}\,^{-1}\coloneqq\frac{2}{4-n}-\gamma+\ln 4\pi$, so one-loop counterterms appear with the universal prefactor $\mathrm{i}/(16\pi^2)\bar{\epsilon}\,^{-1}$. In the symmetric phase the effective operator reads
\beq
\label{eq:B.1}
\cO_{NB,\alpha i}=(\bar{L}_{\alpha}\sigma^{\mu\nu}P_RN_i)\tilde{H}B_{\mu\nu},
\qquad
-\mathcal{L}_{\rm EFT}\supset
C_{NB,\alpha i}(\mu)\,\cO_{NB,\alpha i}+\text{h.c.}
\eeq
Bare and renormalised fields are related by
\beq
\label{eq:B.2}
L_0=Z_L^{1/2}L,\quad
N_0=Z_N^{1/2}N,\quad
H_0=Z_H^{1/2}H,\quad
B_{0\mu}=Z_B^{1/2}B_{\mu},
\eeq
where $Z_X$ denotes the wave-function renormalisation of the field $X$. At one loop we write
\beq
\label{eq:B.3}
Z_X=1+\delta Z_X^{(1)}.
\eeq
Note that, to one loop, the renormalisation of $\CNB$ is multiplicative and diagonal in the sterile-neutrino index $i$.

The counterterm Lagrangian of the $\mathrm{SU}(2)_L$ sector is
\begin{align}
\mathcal{L}_{\rm count.}
&=-\dfrac{1}{2}(Z_3-1)(\partial_{\mu}W_{\nu}^a-\partial_{\nu}W_{\mu}^a)^2
+(Z_2-1)\bar{\psi}\mathrm{i}\gamma^{\mu}P_L\partial_{\mu}\psi\notag\\
&\quad
-(Z_2Z_m-1)m\bar{\psi}\psi
+(\tilde{Z}_3-1)\bar{c}^a\partial_{\mu}\partial^{\mu}c^a\notag\\
&\quad
-(Z_{W^3}-1)gf^{abc}\partial_{\mu}W_{\nu}^aW^{b\mu}W^{c\nu}
-\dfrac{1}{4}(Z_{W^4}-1)g^2(f^{eab}W_{\mu}^aW_{\nu}^b)^2\notag\\
&\quad
+(Z_1Z_2Z_3^{-1}-1)gW_{\mu}^a\bar{\psi}\gamma^{\mu}T^aP_L\psi
+(\tilde{Z}_1-1)g\partial_{\mu}\bar{c}^af^{abc}W_{\mu}^bc^c,
\label{eq:B.4}
\end{align}
with
\beq
\label{eq:B.5}
\begin{split}
W_{0\mu}&=Z_3^{1/2}W_{\mu},\qquad
\psi_0=Z_2^{1/2}\psi,\qquad
(c_0,\bar{c}_0)=\tilde{Z}_3^{1/2}(c,\bar{c}),\\
m_0&=Z_mm,\qquad
g_0=Z_1Z_3^{-3/2}g.
\end{split}
\eeq
For illustrative purposes, the $\mathrm{U}(1)_Y$ sector is not shown here.

\subsection{Representative two-point function: $N$ from $(L,\tilde{H})$ loop}
\label{sec:B.1}

\begin{figure}[t]
\centering
\begin{tikzpicture}[baseline=-2pt,scale=1.5]
  \begin{feynhand}
    \vertex (p1) at (-2.25, 0);
    \vertex (p2) at (2.25, 0);
    \vertex (a) at (-1, 0);
    \vertex (b) at (1, 0);
    \propagator [fer] (a) to [edge label={$q$}] (p1);
    \propagator [fer] (p2) to [edge label={$q$}] (b);
    \propagator [fer] (b) to [edge label={$k$}] (a);
    \propagator [chasca, half left, looseness=1.75] (a) to [edge label={$k-q$}] (b);
  \end{feynhand}
\end{tikzpicture}
\caption{One-loop self-energy of a right-handed (Majorana) neutrino $N_i$ in the symmetric phase. The lepton--Higgs pair $(L,\tilde{H})$ runs in the loop with external momentum $q$. This graph yields the UV pole quoted in eq.~(\ref{eq:B.9}).}
\label{fig:9}
\end{figure}
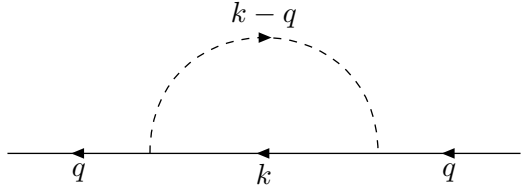

As an explicit example we compute the UV pole of the right-handed (Majorana) neutrino $N_i$ two-point function induced by the lepton--Higgs loop with external momentum $q$. Since $N_i$ is an $\mathrm{SU}(2)_L\times\mathrm{U}(1)_Y$ singlet, there is no contribution from $B$- or $W$-boson loops. The two-point function reads 
\begin{align}
\mathrm{i}\varSigma_{N,\beta\alpha}(q)
&=\int\!\!\dfrac{\dd^nk}{(2\pi)^n}\;
(-\mathrm{i}y_{\beta i}P_R)
\dfrac{\mathrm{i}\slashed{k}}{k^2}
(-\mathrm{i}y_{i\alpha}^{\ast}P_L)
\dfrac{\mathrm{i}}{(k-q)^2}\notag\\
&=(yy^{\dagger})_{\beta\alpha}\int\!\!\dfrac{\dd^nk}{(2\pi)^n}\;
\dfrac{\slashed{k}P_L}{k^2(k-q)^2}.
\label{eq:B.6}
\end{align}
Introducing a Feynman parameter one obtains
\beq
\label{eq:B.7}
\dfrac{1}{k^2(k-q)^2}
=\int_{0}^{1}\dd x\;\dfrac{1}{[(k-q)^2x+k^2(1-x)]^2}
=\int_{0}^{1}\dd x\;
\dfrac{1}{(-q^2x+2k\cdot xq-k^2)^2}.
\eeq
Carrying out the loop integral in dimensional regularisation yields
\begin{align}
\mathrm{i}\varSigma_{N,\beta\alpha}(q)
&=(yy^{\dagger})_{\beta\alpha}
\int_{0}^{1}\dd x\int\!\!\dfrac{\dd^nk}{(2\pi)^n}\;
\dfrac{\slashed{k}P_L}{(-q^2x+2k\cdot xq-k^2)^2}
\notag\\
&=(yy^{\dagger})_{\beta\alpha}\int_{0}^{1}\dd x\;
\dfrac{x\slashed{q}P_L}{[-q^2x(1-x)]^{2-\frac{n}{2}}}\cdot
\dfrac{\mathrm{i}\varGamma\pqty*{2-\frac{n}{2}}}
{(4\pi)^{\frac{n}{2}}\varGamma(2)}\notag\\
&=\dfrac{\mathrm{i}(yy^{\dagger})_{\beta\alpha}}{16\pi^2}
\pqty*{\dfrac{1}{2}\slashed{q}P_L}
\pqty*{\dfrac{2}{4-n}-\gamma+\ln 4\pi}\notag\\
&=\dfrac{\mathrm{i}(yy^{\dagger})_{\beta\alpha}}{16\pi^2}
\bar{\epsilon}\,^{-1}
\pqty*{\dfrac{1}{2}\slashed{q}P_L}.
\label{eq:B.8}
\end{align}
Hence the UV divergence of the amputated two-point function is
\beq
\label{eq:B.9}
\varSigma_{N,\beta\alpha}(q)
=\dfrac{(yy^{\dagger})_{\beta\alpha}}{16\pi^2}\bar{\epsilon}\,^{-1}
\pqty*{\dfrac{1}{2}\slashed{q}P_L},
\eeq
which must be cancelled by the wave-function counterterm from the $\mathrm{SU}(2)_L$ counterterm Lagrangian in eq.~(\ref{eq:B.4}), yielding
\beq
\label{eq:B.10}
\varSigma_N(q)+\delta Z_N^{(1)}\slashed{q}P_L=0
\quad\Rightarrow\quad
\delta Z_{N,\beta\alpha}^{(1)}
=-\dfrac{1}{32\pi^2}\bar{\epsilon}\,^{-1}(yy^{\dagger})_{\beta\alpha}.
\eeq

\subsection{Gauge-boson and matter wave-function counterterms}
\label{sec:B.2}

Collecting the UV poles of the relevant two-point functions in the symmetric phase (all computed in the $\xi=1$ gauge), we obtain
\beq
\label{eq:B.11}
\begin{split}
\delta Z_L^{(1)}
&=-\dfrac{1}{16\pi^2}\bar{\epsilon}\,^{-1}
\bqty*{\pqty*{\dfrac{1}{4}g_1^2+\dfrac{3}{4}g_2^2}\bm{1}
+\dfrac{1}{2}yy^{\dagger}},\quad
\delta Z_N^{(1)}
=-\dfrac{1}{32\pi^2}\bar{\epsilon}\,^{-1}yy^{\dagger},\\[0.5em]
\delta Z_H^{(1)}
&=-\dfrac{1}{16\pi^2}\bar{\epsilon}\,^{-1}
\bqty*{\pqty*{\dfrac{1}{2}g_1^2+\dfrac{3}{2}g_2^2-3y_t^2}\bm{1}
-yy^{\dagger}},\quad
\delta Z_B^{(1)}=-\dfrac{1}{16\pi^2}
\bar{\epsilon}\,^{-1}\cdot\dfrac{41}{6}g_1^2,
\end{split}
\eeq
where $\bm{1}$ denotes the identity matrix in the flavour space of right-handed neutrinos. Converting to anomalous dimensions with~\cite{Kugo1989II, Schwartz_2013}
\beq
\label{eq:B.12}
\gamma_i=\dfrac{1}{2}\pqty*{\mu\pdv{\ln Z_i}{\mu}}_{g_0,\bar{\epsilon}},
\qquad
\pqty*{\mu\dfrac{\partial}{\partial\mu}}_{g_0,\bar{\epsilon}}\ln Z_i
=-2\hbar\pqty*{\dfrac{\partial\delta Z_i^{(1)}}{\partial\bar{\epsilon}\,^{-1}}},
\eeq
gives
\beq
\label{eq:B.13}
\begin{split}
\gamma_L^{(1)}
&=\dfrac{1}{16\pi^2}
\bqty*{\pqty*{\dfrac{1}{4}g_1^2+\dfrac{3}{4}g_2^2}\bm{1}
+\dfrac{1}{2}yy^{\dagger}},\qquad
\gamma_N^{(1)}=\dfrac{1}{32\pi^2}yy^{\dagger},\\[0.5em]
\gamma_H^{(1)}
&=\dfrac{1}{16\pi^2}
\bqty*{\pqty*{\dfrac{1}{2}g_1^2+\dfrac{3}{2}g_2^2-3y_t^2}\bm{1}-yy^{\dagger}},\qquad
\gamma_B^{(1)}=\dfrac{1}{16\pi^2}\cdot\dfrac{41}{6}g_1^2.
\end{split}
\eeq
At one loop the off-diagonal vacuum polarisation $\langle B_{\mu}W_{\nu}^3\rangle$ is proportional to the hypercharge-isospin sum $\sum YT^3$, which vanishes for the complete SM multiplet content (cf. table~\ref{tab:3}). Consequently,
\beq
\label{eq:B.14}
\delta Z_{BW}^{(1)}=0,\qquad
Z_{BW}=1,
\eeq
and there is no $B$--$W^3$ kinetic mixing at this order. 
Using eq.~(\ref{eq:B.2}), renormalisation of $\ONB$ in the symmetric phase follows from
\beq
\label{eq:B.15}
(\bar{L}_{0\alpha}\sigma^{\mu\nu}N_{0i})\tilde{H}_0B_{0\mu\nu}
=(Z_LZ_NZ_HZ_B)^{1/2}
(\bar{L}_{\alpha}\sigma^{\mu\nu}P_RN_i)\tilde{H}B_{\mu\nu},
\eeq
whence the EFT Lagrangian can be written as
\beq
\label{eq:B.16}
-\mathcal{L}_{\rm EFT}\supset
(Z_LZ_NZ_HZ_B)^{1/2}C_{0NB}\cO_{0NB}
\eqqcolon\CNB\ONB,
\eeq
and the renormalisation of $\CNB$ reads
\beq
\label{eq:B.17}
C_{0NB}=Z_{NB}\CNB,\qquad
Z_{NB}\coloneqq (Z_LZ_NZ_HZ_B)^{-1/2}.
\eeq
Therefore, the anomalous dimension of $\ONB$ at one loop is
\begin{align}
\gamma_{NB}
&=\dfrac{1}{2}\pqty*{\mu\pdv{\ln Z_{NB}}{\mu}}_{g_0,\bar{\epsilon}}
=-\dfrac{1}{4}\pqty*{\mu\dfrac{\partial}{\partial\mu}}_{g_0,\bar{\epsilon}}
(\ln Z_L+\ln Z_N+\ln Z_H+\ln Z_B)\notag\\
&=\dfrac{1}{2}
\pqty*{\gamma_L^{(1)}+\gamma_N^{(1)}+\gamma_H^{(1)}+\gamma_B^{(1)}}
=\dfrac{1}{32\pi^2}
\pqty*{\dfrac{91}{12}g_1^2+\dfrac{9}{4}g_2^2-3y_t^2}\bm{1}.
\label{eq:B.18}
\end{align}
The $yy^{\dagger}$ terms cancel exactly between field renormalisations, so $\gamma_{NB}$ contains only gauge and top-Yukawa pieces.
Differentiating $C_{0NB}=Z_{NB}\CNB$ with respect to $\mu$ gives
\beq
\label{eq:B.19}
\mu\dfrac{\dd}{\dd\mu}Z_{NB}=2\gamma_{NB}Z_{NB}
\quad\Rightarrow\quad
\mu\dfrac{\dd}{\dd\mu}\CNB=-2\gamma_{NB}\CNB.
\eeq
Substituting~(\ref{eq:B.18}) into the latter equation of~(\ref{eq:B.19}) leads to the renormalisation-group equation
\beq
\label{eq:B.20}
\mu\dfrac{\dd}{\dd\mu}\CNB
=-\dfrac{1}{16\pi^2}
\pqty*{\dfrac{91}{12}g_1^2+\dfrac{9}{4}g_2^2-3y_t^2}\CNB,
\eeq
which is consistent with~(\ref{eq:B.14}); the $B$--$W^3$ kinetic mixing does not enter at this order.

\section{Computation of decay widths and CP asymmetries}
\label{sec:C}

In this appendix, we derive explicitly the two-body decay widths for the photon and the $Z$ modes, and we compute the CP asymmetries in detail only for the photon mode to keep the presentation concise. The $Z$-mode asymmetries follow from the broken-phase relation in eq.~(\ref{eq:4.15}). A step-by-step evaluation of the one-loop topologies in a closely related setup can be found in ref.~\cite{Law:2008yyq}.

\subsection{Tree-level width for the photon mode}
\label{sec:C.1}

In the broken phase the interaction relevant for $N_i\to\nu_{\alpha}\gamma$ reads
\beq
\label{eq:C.1}
-\mathcal{L}_{\rm EFT}\supset
\mu_{\alpha i}(\bar{L}_{\alpha}\sigma^{\mu\nu}P_RN_i)F_{\mu\nu}
+\text{h.c.},\qquad
\mu_{\alpha i}=\dfrac{v\cos\theta_{\rm W}}{\sqrt{2}\,M_{\Psi}^2}\,
C_{NB,\alpha i}(\mu_{\rm ref}),
\eeq
where $\sigma^{\mu\nu}\coloneqq
\frac{\mathrm{i}}{2}[\gamma^{\mu},\gamma^{\nu}]$ and $F_{\mu\nu}=\partial_{\mu}A_{\nu}-\partial_{\nu}A_{\mu}$ is the photon field strength.

For an outgoing photon of momentum $q$ and polarisation $\varepsilon_{\rho}^{\ast}(q)$ the Feynman rule reads
\beq
\label{eq:C.2}
V_{\alpha i}^{\rho}=2\mu_{\alpha i}\sigma^{\rho\lambda}q_{\lambda}P_R.
\eeq
The tree-level amplitude for $N_i(k)\to\nu_{\alpha}(p)+\gamma_{\rho}(q)$ is therefore
\beq
\label{eq:C.3}
\mathrm{i}\mathcal{M}_{\gamma}^{\rm tree}
=\mu_{\alpha i}\bar{u}_{\alpha}(p)(2\sigma^{\rho\lambda}q_{\lambda}P_R)
u_i(k)\varepsilon_{\rho}^{\ast}(q),
\eeq
with $k=p+q$. Setting $m_{\nu}=0$ (hence $p^2=0$) and summing over spins and polarisations one finds
\beq
\label{eq:C.4}
\overline{\abs{\mathcal{M}_{\gamma}^{\rm tree}}^2}
=-2\abs{\mu_{\alpha i}}^2\tr[\sigma^{\rho\lambda}q_{\lambda}P_R(\slashed{k}+M_i)P_L{\sigma_{\rho}}^{\beta}q_{\beta}].
\eeq
Using
\beq
\label{eq:C.5}
\sigma^{\mu\nu}q_{\nu}
=\dfrac{\mathrm{i}}{2}(\gamma^{\mu}\slashed{q}-\slashed{q}\gamma^{\mu}),
\qquad
\sigma^{\mu\nu}\gamma^{\rho}
=\mathrm{i}(g^{\nu\rho}\gamma^{\mu}-g^{\mu\rho}\gamma^{\nu})
-\epsilon^{\mu\nu\rho\sigma}\gamma_{\sigma}\gamma_5,\quad
(\epsilon^{0123}=+1),
\eeq
and adopting the kinematics $k\cdot q=p\cdot q=M_i^2/2$, we obtain
\beq
\label{eq:C.6}
\overline{\abs{\mathcal{M}_{\gamma}^{\rm tree}}^2}
=4\abs{\mu_{\alpha i}}^2M_i^4.
\eeq
The two-body phase space then gives~\cite{PhysRevD.78.085024, Law:2008yyq}
\beq
\label{eq:C.7}
\varGamma_{\gamma,\alpha i}^{\rm tree}
=\dfrac{\abs{\mu_{\alpha i}}^2M_i^3}{2\pi}
=\dfrac{v^2\cos^2\theta_{\rm W}}{4\pi}\dfrac{M_i^3}{M_{\Psi}^4}
\abs{C_{NB,\alpha i}(\mu_{\rm ref})}^2,\qquad
\varGamma_{\gamma,i}^{\rm tree}
=\sum_{\alpha}\varGamma_{\gamma,\alpha i}^{\rm tree}.
\eeq

\subsection{Tree-level width for the $Z$ mode}
\label{sec:C.2}

In the broken phase $A_{\mu}=B_{\mu}\cos\theta_{\rm W}$ and $Z_{\mu}=-B_{\mu}\sin\theta_{\rm W}$. Hence, the dipole that mediates $N_i\to\nu_{\alpha}Z$ is related to the electromagnetic one by
\beq
\label{eq:C.8}
\mu_{\alpha i}^{Z}=-\mu_{\alpha i}\tan\theta_{\rm W}.
\eeq
The matrix element has the same Lorentz structure as in the photon case, and for a massive vector the polarisation sum is
\beq
\label{eq:C.9}
\sum_{\rm pol.}\varepsilon_{\mu}^{\ast}(q)\varepsilon_{\nu}(q)
=-g_{\mu\nu}+\dfrac{q_{\mu}q_{\nu}}{m_Z^2}.
\eeq
The longitudinal term drops because $q_{\mu}\sigma^{\mu\nu}q_{\nu}=0$. Using $q^2=m_Z^2$ and
\beq
\label{eq:C.10}
p\cdot q=\dfrac{M_i^2-m_Z^2}{2},\qquad
q\cdot k=\dfrac{M_i^2+m_Z^2}{2},
\eeq
we find
\beq
\label{eq:C.11}
\overline{\abs{\mathcal{M}_Z^{\rm tree}}^2}
=4\abs{\mu_{\alpha i}}^2(M_i^4-m_Z^4)\tan^2\theta_{\rm W}.
\eeq
Performing the two-body phase-space integral,
\beq
\label{eq:C.12}
\varGamma_{Z,\alpha i}^{\rm tree}
=\dfrac{\abs{\mu_{\alpha i}}^2M_i^3}{2\pi}
\pqty*{1-\dfrac{m_Z^2}{M_i^2}}^2\pqty*{1+\dfrac{m_Z^2}{M_i^2}}
\tan^2\theta_{\rm W}.
\eeq
Substituting $\mu_{\alpha i}=\frac{v\cos\theta_{\rm W}}{\sqrt{2}\,M_{\Psi}^2}\,C_{NB,\alpha i}(\mu_{\rm ref})$ yields
\beq
\label{eq:C.13}
\varGamma_{Z,\alpha i}^{\rm tree}
=\dfrac{v^2\sin^2\theta_{\rm W}}{4\pi}\dfrac{M_i^3}{M_{\Psi}^4}
\pqty*{1-\dfrac{m_Z^2}{M_i^2}}^2\pqty*{1+\dfrac{m_Z^2}{M_i^2}}
\abs{C_{NB,\alpha i}(\mu_{\rm ref})}^2,\quad
\varGamma_{Z,i}^{\rm tree}
=\sum_{\alpha}\varGamma_{Z,\alpha i}^{\rm tree}.
\eeq
Finally, the ratio to the photon width of sec.~\ref{sec:C.1} is
\beq
\label{eq:C.14}
\dfrac{\varGamma_Z^{\rm tree}}{\varGamma_{\gamma}^{\rm tree}}
=(1-r_Z)^2(1+r_Z)\tan^2\theta_{\rm W},\qquad
r_Z\coloneqq\dfrac{m_Z^2}{M_i^2},
\eeq
so that $\varGamma_Z^{\rm tree}/\varGamma_{\gamma}^{\rm tree}\to\tan^2\theta_{\rm W}$ for $M_i\gg m_Z$.

\subsection{Ingredients for CP asymmetry}
\label{sec:C.3}

\subsubsection{Optical theorem and Cutkosky rules}
\label{sec:C.3.1}

The CP asymmetry originates from the imaginary part of the interference between the tree and one-loop amplitudes. Rather than computing imaginary parts directly, we use the optical theorem
\beq
\label{eq:C.15}
2\Im\mathcal{M}_{fi}
=\sum_{n}\int\dd\Pi_n\;\mathcal{M}_{fn}\mathcal{M}_{in}^{\ast},
\eeq
which implies that the discontinuity across a physical cut is
\beq
\label{eq:C.16}
\Disc(\mathcal{M})=2\mathrm{i}\Im\mathcal{M}.
\eeq
For each cut we replace the cut propagators by their on-shell delta function,
\beq
\label{eq:C.17}
\dfrac{1}{p^2-m^2+\mathrm{i}\epsilon}
\quad\longrightarrow\quad
-2\pi\mathrm{i}\,\delta(p^2-m^2)\Theta(\pm E).
\eeq
The $\Theta(\pm E)$ selects the positive-/negative-energy branch along the cut. Eq.~(\ref{eq:C.17}) follows the distributional identity
\beq
\label{eq:C.18}
\dfrac{1}{x\pm\mathrm{i}\epsilon}
=\mathcal{P}\pqty*{\dfrac{1}{x}}
\mp\mathrm{i}\pi\delta(x).
\eeq
In practice, we (i) list all kinematically allowed cuts that put the internal lines on shell simultaneously; (ii) replace the cut propagators as in eq.~(\ref{eq:C.17}) with the appropriate $\Theta$-functions; (iii) perform the loop integrals; and (iv) sum all cuts and use eq.~(\ref{eq:C.16}) to reconstruct $\Im\mathcal{M}$. This reduces the problem to phase-space integrals of on-shell amplitudes~\cite{10.1063/1.1703676}.

\subsubsection{Denner's fermion-flow prescription (Majorana lines)}
\label{sec:C.3.2}

We treat Majorana fermions using Denner's rules~\cite{DENNER1992467} and do not rewrite amplitudes by explicit charge conjugation matrices. The reversed spinor structure is defined by
\beq
\label{eq:C.19}
\Gamma^{\rm rev}\coloneqq C\Gamma^{\intercal}C^{-1}
\eqqcolon\eta_{\Gamma}\,\Gamma,\qquad
\eta_{\Gamma}=\begin{cases}
+1,&\Gamma=1,\;\mathrm{i}\gamma_5,\;\gamma^{\mu}\gamma_5,\\
-1,&\Gamma=\gamma^{\mu},\;\sigma^{\mu\nu}.
\end{cases}
\eeq
The following are the working rules for applying Denner's fermion-flow prescription:
\begin{enumerate}
\item Enumerate graphs. Draw all Feynman graphs contributing to the process.

\item Fix a fermion flow. For each fermion chain, assign an arbitrary but fixed orientation (fermion flow) and mark the arrows on the graph.

\item Read spinor strings. Start from an external leg (for closed loops from an arbitrary propagator) and write down the Dirac string opposite to the chosen flow along the chain, so that the spinor chain has the standard left-to-right form $\bar{u}(\cdots)u$.

\item Insert building blocks with a local test. At each internal propagator, external leg and vertex, compare the reading direction (left $\to$ right in the chain you are writing) with the local fermion-flow arrow at that element (as drawn in the graph):
\begin{itemize}
\item if they are antiparallel (reading opposite to the arrow), use the standard objects $\Gamma$ and $S(p)$, where $S(p)$ denotes the fermion propagator;
\item if they are parallel (reading along the arrow), replace by the reversed objects
\beq
\label{eq:C.20}
\Gamma\;\longrightarrow\;\Gamma^{\rm rev}
=\eta_{\Gamma}\,\Gamma,\qquad
S(p)\;\longrightarrow\;S(-p),
\eeq
with $\eta_{\Gamma}$ as in eq.~(\ref{eq:C.19}).
\end{itemize}
\item Operational fact. The local arrow at a vertex is taken from the graph once the flow is fixed.
\end{enumerate}
These prescriptions uniquely reproduce the signs and spinor structures used in sec.~\ref{sec:C.4} and in the main text.



\subsubsection{Double Majorana back-flow}
\label{sec:C.3.3}

Electromagnetic leptogenesis is mediated by
\beq
\label{eq:C.21}
\cO_{NB,\alpha i}
=(\bar{L}_{\alpha}\sigma^{\mu\nu}P_RN_i)\tilde{H}B_{\mu\nu}.
\eeq
Since the operator couples the neutrino and antineutrino components, each loop graph admits two on-shell cuts that differ only by the direction of the lepton-number flow through the Majorana line. Denoting schematically
\beq
\label{eq:C.22}
\mathcal{M}_{\rm loop}^{\rm (p)}\propto
\mu_{\alpha m}\mu_{\beta m}\mu_{\beta i}^{\ast},\qquad
\mathcal{M}_{\rm loop}^{\rm (ap)}\propto
\mu_{\alpha m}^{\ast}\mu_{\beta m}^{\ast}\mu_{\beta i},
\eeq
the two interferences with the tree amplitude contribute equally to $\Im\mathcal{M}$. Thus, the CP-odd piece is doubled,
\beq
\label{eq:C.23}
\begin{split}
\Im(A_{\mu}^{(5)})
&=\Im(A_{\mu}^{(5)})^{\rm (p)}
-\Im(A_{\mu}^{(5)})^{\rm (ap)}\\
&=\Im(A_{\mu}^{(5)})^{\rm (p)}
-\Im(A_{\mu}^{(5)\ast})^{\rm (p)}\\
&=2\Im(A_{\mu}^{(5)})^{\rm (p)},
\end{split}
\eeq
where $A_{\mu}^{(5)}$ is defined in eq.~(\ref{eq:C.30}). The tree amplitude is linear in $\mu_{\alpha i}$, while each loop cut carries three dipole couplings; under conjugation, the CP-odd interference flips sign, so the two cuts add in $\Im\mathcal{M}$. This doubling is absent for Dirac neutrinos.


\subsection{Illustrative calculation of CP asymmetries}
\label{sec:C.4}

\enlargethispage{\baselineskip}

The tree amplitude for $N_i(k)\to\nu_{\alpha}(p)+\gamma_{\rho}(q)$ is given by
\beq
\label{eq:C.24}
\mathrm{i}\mathcal{M}_{\gamma}^{\rm tree}
=2\mu_{\alpha i}\bar{u}_{\alpha}(p)\sigma^{\rho\lambda}q_{\lambda}
P_Ru_i(k)\varepsilon_{\rho}^{\ast}(q).
\eeq
For the vertex graph of fig.~\ref{fig:4}(a) the loop amplitude contains the propagators
\beq
\label{eq:C.25}
S_{\nu}(q_1)\simeq\dfrac{\mathrm{i}\slashed{q}_1}{q_1^2+\mathrm{i}\epsilon},
\qquad
S_{N_m}(q_3)=\dfrac{\mathrm{i}(\slashed{q}_3+M_m)}{q_3^2-M_m^2+\mathrm{i}\epsilon},\qquad
D_{\mu\nu}(q_2)=\dfrac{-\mathrm{i}g_{\mu\nu}}{q_2^2+\mathrm{i}\epsilon},
\eeq
and the effective vertices
\beq
\label{eq:C.26}
\begin{split}
\tilde{V}_{\beta i}^{\rho}(q_2)
&=2\mu_{\beta i}^{\ast}\sigma^{\rho\lambda}q_{2\lambda}P_L,\\
V_{\alpha m}^{\mu}(-q_2)
&=2\mu_{\alpha m}\sigma^{\mu\eta}(-q_{2\eta})P_R,\\
V_{\beta m}^{\sigma}(q)
&=2\mu_{\beta m}\sigma^{\sigma\kappa}q_{\kappa}P_R.
\end{split}
\eeq
The CP-odd contribution arises from the interference,
\beq
\label{eq:C.27}
I_{\rm vert}^{\rm (a)}\coloneqq
\dfrac{1}{2}\sum_{\rm spins}\sum_{\rm pol.}
\pqty*{\mathrm{i}\mathcal{M}_{\gamma}^{\rm tree}}^{\dagger}
\pqty*{\mathrm{i}\mathcal{M}_{\gamma}^{\text{\scriptsize 1-loop}}}
\eeq
Fixing the fermion-flow as $u_i\to\bar{u}_{\alpha}$ and writing the Dirac string opposite to it (Denner's rules 1.--3.), the chain $\bar{u}_{\alpha}V_{\alpha m}^{\mu}S_{N_m}V_{\beta m}^{\sigma}S_{\nu}\tilde{V}_{\beta i}^{\rho}u_i$ is read left-to-right antiparallel to the local arrows taken from the graph in fig.~\ref{fig:4}(a). Hence, the local test triggers no replacements in eq.~(\ref{eq:C.20}), and no additional minus sign is generated~\cite{DENNER1992467}.
Using $\gamma^0\sigma^{\mu\nu}\gamma^0=-\sigma^{\mu\nu}$ one finds
\beq
\label{eq:C.28}
\pqty*{\mathrm{i}\mathcal{M}_{\gamma}^{\rm tree}}^{\dagger}
=2\mu_{\alpha i}^{\ast}\varepsilon_{\rho}(q)\bar{u}_i(k)P_L\sigma^{\rho\lambda}q_{\lambda}u_{\alpha}(p).
\eeq
After the photon polarisation sum $\sum_{\rm pol.}\varepsilon_{\rho}^{\ast}\varepsilon_{\sigma}\to -g_{\rho\sigma}$, the spin sums and the Dirac trace, the interference integral becomes
\begin{align}
I_{\rm vert}^{\rm (a)}
&=\mathrm{i}A_{\mu}^{(5)}\int\dfrac{\dd^4q_1}{(2\pi)^4}\;
\dfrac{M_iM_m}
{(q_3^2-M_m^2+\mathrm{i}\epsilon)(q_1^2+\mathrm{i}\epsilon)
(q_2^2+\mathrm{i}\epsilon)}\notag\\
&\quad\times
\bigl\{-32M_i^2(q\cdot q_1)q_2^2\notag\\
&\hspace{2.5em}
+64(q\cdot q_2)\bqty*{-2(p\cdot q_1)(q\cdot q_2)+2(p\cdot q_2)(q\cdot q_1)+M_i^2(q_1\cdot q_2)}\bigr\}
\label{eq:C.29}
\end{align}
with
\beq
\label{eq:C.30}
A_{\mu}^{(5)}\coloneqq\mu_{\alpha i}^{\ast}\mu_{\alpha m}\mu_{\beta m}\mu_{\beta i}^{\ast},\qquad
J_{\rm vert}^{(a)}\coloneqq I_{\rm vert}^{(a)}/A_{\mu}^{(5)}.
\eeq
The imaginary part arises from cutting the massless lepton and photon lines.
The Cutkosky replacements,
\beq
\label{eq:C.31}
\dfrac{1}{q_1^2+\mathrm{i}\epsilon}
\to -2\pi\mathrm{i}\,\delta(q_1^2)\Theta(E_1),\qquad
\dfrac{1}{q_2^2+\mathrm{i}\epsilon}
\to -2\pi\mathrm{i}\,\delta(q_2^2)\Theta(E_2),
\eeq
and used in the $N_i$ rest frame,
\beq
\label{eq:C.32}
\begin{split}
&k=(M_i,\bm{0}),\qquad
p=\pqty*{\dfrac{M_i}{2},-\bm{q}},\qquad
q=\pqty*{\dfrac{M_i}{2},\bm{q}},\qquad
\norm{\bm{q}}=\dfrac{M_i}{2},\\
&k^2=M_i^2,\qquad
p^2=0,\qquad
q^2=0,\qquad
q_2=k-q_1,\qquad
q_3=q_1-q.
\end{split}
\eeq
Performing the angular integrals, the discontinuity of eq.~(\ref{eq:C.29}) takes the compact form
\begin{align}
\Disc(J_{\rm vert}^{\rm (a)})
&=\dfrac{\mathrm{i}}{4\pi}M_i^5M_m
\int_{-1}^{1}\dd t\;\dfrac{(1+t)\bqty*{-(1+t)^2+(1-t)^2+4}}{1-t+2x}\notag\\
&=\dfrac{2\mathrm{i}}{\pi}M_i^6f_{V_a}(x),\qquad
x\coloneqq\dfrac{M_m^2}{M_i^2},
\label{eq:C.33}
\end{align}
which defines the real loop function $f_{V_a}(x)$ used in the main text,
\beq
\label{eq:C.34}
f_{V_a}(x)=\sqrt{x}\,\Bqty*{1+2x\bqty*{1-(1+x)\ln\dfrac{1+x}{x}}}.
\eeq
Including the double-flow factor in eq.~(\ref{eq:C.23}), the vertex contribution to the photon-mode CP asymmetries from fig.~\ref{fig:4}(a) is
\begin{align}
\varepsilon_{{\rm vert},\alpha i}^{\gamma,{\rm (a)}}
&=-\dfrac{4}{\varGamma_{\gamma, i}^{\rm tree}}
\Im(A_{\mu}^{(5)})\Im(J_{\rm vert}^{\rm (a)}\varPhi_2)\notag\\
&=-\dfrac{M_i^2}{2\pi\sum_{\beta}\abs{\mu_{\beta i}}^2}\sum_{m\neq i}
\Im\bqty*{\mu_{\alpha i}^{\ast}\mu_{\alpha m}(\mu^{\dagger}\mu)_{im}}\,
f_{V_a}\!\pqty*{\dfrac{M_m^2}{M_i^2}},
\label{eq:C.35}
\end{align}
where $\varPhi_2=\norm{\bm{q}}/(8\pi M_i^2)$ denotes the two-body phase-space factor.
Eq.~(\ref{eq:C.35}) is gauge invariant.

The interference for fig.~\ref{fig:4}(b) vanishes after the momentum-conservation identities and $q^2=0$ are applied, while the self-energy graphs reproduce the loop functions $f_{S_a}(x)$, $f_{S_b}(x)$ in eq.~(\ref{eq:4.14}). Collecting all contributions yields the photon-mode asymmetries in eq.~(\ref{eq:4.13}); the $Z$-mode results follow from eq.~(\ref{eq:4.15}). The total CP asymmetries are shown in eq.~(\ref{eq:4.16}).

\bibliographystyle{JHEP}
\bibliography{refs}
\end{document}